# A Multi-Vocal Review of Security Orchestration

CHADNI ISLAM, University of Adelaide, CREST – The Centre for Research on Engineering Software Technologies, and Data61, CSIRO, Australia
MUHAMMAD ALI BABAR, University of Adelaide and CREST – The Centre for Research on Engineering Software Technologies, Australia
SURYA NEPAL, Data61, CSIRO, Australia

## Abstract

Organizations use diverse types of security solutions to prevent cyber-attacks. Multiple vendors provide security solutions developed using heterogeneous technologies and paradigms. Hence, it is a challenging rather impossible to easily make security solutions to work an integrated fashion. Security orchestration aims at smoothly integrating multivendor security tools that can effectively and efficiently interoperate to support security staff of a Security Operation Centre (SOC). Given the increasing role and importance of security orchestration, there has been an increasing amount of literature on different aspects of security orchestration solutions. However, there has been no effort to systematically review and analyze the reported solutions. We report a Multivocal Literature Review that has systematically selected and reviewed both academic and grey (blogs, web pages, white papers) literature on different aspects of security orchestration published from January 2007 until July 2017. The review has enabled us to provide a working definition of security orchestration and classify the main functionalities of security orchestration into three main areas – unification, orchestration, and automation. We have also identified the core components of a security orchestration platform and categorized the drivers of security orchestration based on technical and socio-technical aspects. We also provide a taxonomy of security orchestration based on the execution environment, automation strategy, deployment type, mode of task and resource type. This review has helped us to reveal several areas of further research and development in security orchestration.

## Additional Key Words and Phrase

Security orchestration, security automation, multivocal literature review, intelligent security assistant.

## 1. Introduction

The cybersecurity breaches lead to serious organizational and socio-economic consequences such as loss of revenue, damage to reputation and information system, theft of proprietary data and customer sensitive information [1-4]. For example, Equifax (one of the largest credit reporting agencies in America) [5] reported a major data breach that had affected around 148 million US consumers [6-8]. The hackers successfully stole sensitive information (e.g., credit card number, phone number, email address, and social security number) through that breach, which was preventable as per a recent report. According to a research sponsored by IBM, the average total cost of a breach is around $3.62 million per incident [4].

Organizations are using various security solutions to prevent known and unknown attacks and avoid the consequences usually associated with security vulnerabilities and threats [9, 10]. Some of the commonly used security solutions are antivirus, Firewall, Intrusion Detection Systems and Instruction Prevention Systems (IDS/IPS), and Security Information and Events Management (SIEM) [9, 11-13]. The security solutions providers use different technologies and paradigms to develop, deploy, and operate their security solutions, which cannot be easily integrated and interoperated for effectively and efficiently supporting Security Operation Centre (SOC).

Security orchestration is aimed at introducing technical and socio-technical solutions to integrate multivendor security tools as a unified whole to support security staff in a SOC. Organizations are increasingly adopting security orchestration platforms that are proactive, autonomous, and collaborative solutions to enable security staff perform their responsibilities effectively and efficiently [14-17]. A security orchestration initiative enables peoples, practices, and technologies to work together to improve organizations' security intelligence for better security operations and management [18-20]. Security orchestration is a prerequisite of security automation, which is the process of automatically detecting, preventing, and recovering from cyber-attacks without human interference using information technology, automation algorithm, and artificial intelligence [19, 21].

Existing security solutions are designed to monitor an organization's IT infrastructures and network activities, generate security alerts, and perform necessary actions upon detection of security threats. An organization's cybersecurity solutions generate thousands of alerts, which are usually monitored and acted upon by security staff, mostly using manual or semi-automated processes and practices [21-23]. A Verizon's report indicates that 93% of the data breach cases require minutes to be executed, but it took companies weeks or months to discover the attacks [24]. For example, after getting alerts from IDS for malicious behaviors, a security expert might go to an endpoint defense system to gather more relevant information by querying network resources and validating the threat. After confirming the threat, a security expert commands a Firewall to isolate or block the traffic from the affected region and update the threat information in the threat intelligence. According to BakerHostetler [25], security experts took on average 61 days to discover the occurrence of an incident and after discoveries 41 more days to take remedial actions. A food chain, Wendy's Point of Sale systems were affected by malware at 1025 locations in 2015, but it was first discovered in February 2016 [26, 27]. To deal with the potential threats to security breaches, security experts are expected to provision and facilitate the selection of existing security solutions as quickly as possible to provide the required security services and ensure seamless security operation.

A security orchestration solution has the potential to address the concerns of manual threat analysis, delays in responses to security incidents as well as provide security status of an organization's ICT infrastructure. Security orchestration solutions are capable of automatically identify suspicious activities in an organization environment, and proactively act to mitigate a cyber-attack. According to a Gartner's report, by 2019 30% of the large and medium enterprises will be deploying some forms of security automation and orchestration capabilities [28]. Another study [23] reports one third of organizations are planning to deploy or have deployed security orchestration solutions that can bundle different security solutions and human expertise together for the automation of security services within an organization.

Fig. 1 captures some of the abovementioned organizational settings where several types of security solutions generate alerts to be manually analyzed by security staff in the absence of a security orchestration platform that can automate most of the manual decision-making process against a threat incident. Security orchestration platform integrates security tools to accelerate incident response by reducing the manual and repetitive activities. Orchestrating and automating the activities of multivendor security solutions require a comprehensive view of the orchestration platform as these solutions have their own way to work and produce different formats of alerts. The existing security orchestration solutions do not provide sufficient evidence for supporting different quality attributes such as flexibility, interoperability, scalability, modifiability, accuracy, and extensibility [16, 17, 29-33]. Given the increasing demand for security orchestration, a significant amount of research is needed to help understand the challenges in security orchestration, existing solutions, and practices to address the challenges.

This paper reports a Multi-Vocal Literature Review (MLR) that aimed at systematically identifying and reviewing the literature on the security orchestration "state of the art" and "state of the practice". An MLR (i.e., a type of Systematic Literature Review) includes both peer reviewed and non-peer reviewed literature (e.g., newsletters, white papers, fact sheets, and blog posts) [34, 35]. Systematic Literature Review (SLR) has become the most popular methods of conducting a literature review in Software Engineering (SE) [36].

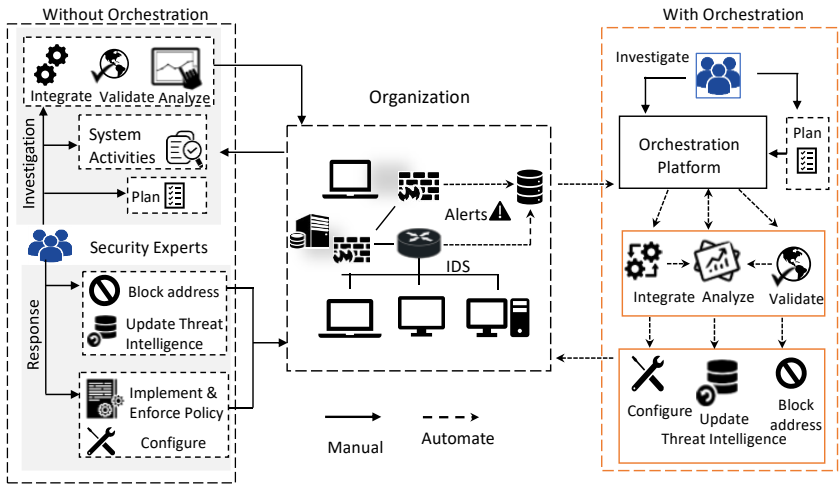

**Fig. 1** Overview of an organization decision against alerts without security orchestration and with security orchestration

SLR focuses only on research contribution and does not include grey literature. SLR cannot always provide an established discipline of knowledge as it ignores a large amount of information produced by software engineering practitioners [34, 35, 37 ]. Hence, MLR is attracting more attention to SE [34, 38]. We believe that conducting MLR in the area of security orchestration will be more useful compared to SLR as there is a large body of non-peer reviewed literature reported by practitioners. This MLR has explored the fundamental challenges and opportunities for the evolution of security orchestration. We have analyzed the characteristics of the existing security orchestration technologies to understand how to solve the security orchestration challenges. We have also investigated the strength and weakness of the reported security orchestration technologies. The main contributions of this MLR are:

- It introduces a working definition of security orchestration, followed by several functionalities of security orchestration ranging from integrating several security solutions to performing incident response planning against a threat as well as collaborative security solutions to materialize the concept of security orchestration (refer to sections 3 & 4 for further details)
- It identifies the key challenges that practitioners and researchers intend to overcome through security orchestration (details are discussed in section 5).
- It provides a taxonomy of different aspect of security orchestration practices needed to support dynamic adoption of applications within the organization environment (detail in section 6).
- It determines and discusses the open research challenges/issues in the field of security orchestration (refer to sections 7 & 8).

## 2. Research Method

The methodology used for this review has benefited from the SLR guideline reported in [39]; there are no specific guidelines available for conducting MLR in SE. Hence, our study's methodology was inspired by the work reported in [40, 41]. The study methodology involves three main phases: planning and designing the review protocol, conducting the review, and reporting the review. We developed a review protocol describing each step of MLR. The review protocol includes several steps: research identification, search strategies, study selection, data extraction, and synthesis. Our MLR process followed the steps in the same order as shown in Fig. 2.

### 2.1. Research Identification

We identified the relevant literature using a search strategy that was based on a set of research questions, as shown in Table 1. The research questions purported to help gain an understanding of security orchestration, the functions and their respective elements of security orchestration. The research questions also aimed at helping in identifying and reviewing the supportive tools, approaches, evaluation criteria for adopting security orchestration in practice.

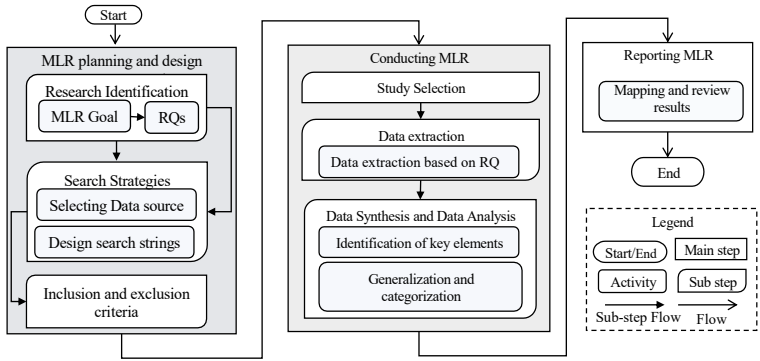

**Fig. 2** An overview of our MLR process

**Table 1** Research Questions of this MLR

| Research Questions | Motivation |
|---|---|
| **RQ1.** What is Security Orchestration? | The first question (RQ1) investigates how security orchestration is defined. RQ1 aims at identifying the relevant related work, i.e., identifying keywords for literature search that leads to a maximum coverage of the related approaches. RQ1 also investigates the functional features and core elements of security orchestration. |
| **RQ2.** What challenges security orchestration intend to solve? | Security orchestration is commonly used by practitioners to bring automation, streamline incident response, and integrate security tools. Many challenges that exist for security solutions in a more traditional setting also apply to security orchestration. RQ2 focuses on the aspects where security orchestration fundamentally differs compared to traditional approaches. |
| **RQ3.** What types of solutions have been proposed? *RQ3.1.* What practices have been reported for adopting security orchestration? *RQ3.2.* What types of tools and techniques researchers and practitioners use, propose, design, and implement in practice? *RQ3.3.* What aspects of architecture security practitioners consider for large-scale deployment of security orchestration? | The motive behind this question is to identify the solutions related to security orchestration and the reported practices followed by organizations (i.e., requirements, guidelines, and collaborative approaches) for adopting security orchestration (RQ3.1), more specifically how existing tools are employed to implement; what are the innovative approaches and technique needed for successful implementation of security orchestration (RQ3.2); and most importantly what aspects of architecture are being considered for large-scale deployment of security orchestration (RQ3.3). RQ3 would help enable researchers to find the gap and practitioners to consider the architectural aspects that need to be considered to successfully implement security orchestration on a larger scale. |

## 2.2. Search Strategy

Following sections detail the search strategy for acquiring the relevant literature from multiple sources.

*2.2.1. Data Sources.* Our review includes both peer reviewed and grey literature that was identified and acquired using both manual and automatic searches in the relevant sources. Initially, we performed a manual search on the *Journal of Computer Security*, *ACM SIGSAC Conference on Computer and Communications Security*, and *USENIX Security Symposium* to gain an overview of the recent literature. We also searched the recent proceedings of RSA conferences. Then we conducted the automatic search in three digital libraries, *IEEE Xplore*, *ACM Digital Library*, and *Scopus*, that publish peer reviewed literature on computing. We used the advanced search option to facilitate the search that allows multiple keywords search. During the automatic search in the digital libraries, we defined the search to match the search string with the titles, abstracts, and keywords of the papers published between January 2007 and July 2017. Our search in *ACM DL, IEEE Xplore*, and *Scopus* include the paper from Annual Computer Security Applications Conference and IEEE Security and Privacy. We additionally searched in Google Scholar to search and include some relevant literature, especially some patents that we could have missed through the abovementioned search procedure.

To search the grey literature, we used the Google search engine like other MLRs [37, 38]. We search the first ten pages, which was considered sufficient to find the most relevant literature as Google search engine's algorithm retrieves and shows the most relevant results in the first few pages [37, 38]. For example, Google search engine returned 45,900 results for the term "security orchestration" in November 2017; however, the relevant content is captured in the first ten pages.

*2.2.2.  Search Strings.*  We created a search string to ensure a thorough search over several databases. For academic literature, we formulated the search string based on a) the key terms gathered from the relevant papers, b) synonym, alternative terms, and related concepts of security orchestration, c) **AND** and **OR** to combine all the terms. We performed several pilot searches and refined, discarded, and added search terms to confirm the inclusion of the relevant papers that we already knew. We formulated the search string in three parts. We performed a match of the search string with the paper's titles, abstracts, and keywords. We used the following search string.

| Search String 1 | *("Security" **OR** "Alert" **OR** "Threat" **OR** "Policy" **OR** "Intrusion" **OR** "Anomaly Detection" **OR** "forensic")* **AND** *("Orchestration" **OR** "Instrumentation" **OR** "Coordination" **OR** "Correlation" **OR** "Collaboration" **OR** "Automation" **OR** "Integration")* **AND** *("Security Tool" **OR** "Safeguard Software" **OR** "IDS" **OR** "IPS" **OR** "Threat Intelligence" **OR** "Detection Engine" **OR** "Prevention Engine" **OR** "Security Control" **OR** "Security Appliance" **OR** "perimeter defense" **OR** "Incident Response")* |
|---|---|

We used the search string *"(Security AND Orchestration)"* to search grey literature and conducted a search on the Google search engine and Google Scholar.

## 2.3. Eligibility Criteria

We defined a set of inclusion and exclusion criteria to select the papers. The criteria are shown in Table 2. Since this review is a blend of scientific and grey literature, we used a rather narrow inclusion and exclusion criteria.

**Table 2 Inclusion and exclusion criteria**

| Inclusion criteria | Exclusion criteria |
|---|---|
| IC1: Articles in English and full text is accessible. | EC1: Short academic paper (paper less than 6 pages). |
| IC2: Articles that focus on developing integrated, coordinated and collaborative solutions. | EC3: Article whose focus is irrelevant to security. |
| IC3: Articles include a sound validation (for grey source: working prototype or tools, proper references to validate the result) | EC4: Article that focuses on physical infrastructure or hardware. |
| IC4: Articles that reports practices and challenges in cyberspace (such as blogs, magazine reports) to give an indication towards orchestration. | EC5: Article that focuses to enhance algorithms or features of a single security solution. |
|  | EC6: All duplicate articles found from various sources. |

## 2.4. Study Selection

Fig. 3 shows the details of the selection of grey and academic literature at each step of this MLR. This also includes the search databases and a number of papers selected after each step. We followed two different approaches to selecting the academic and grey literature.

*2.4.1.Selection of Academic Literature.*  In this section, we describe each step of the process of selecting the relevant papers. Our search in *ACM DL, IEEE Xplore*, and *Scopus* returned 271, 600 and 1017 results, respectively. The titles, abstracts, and keywords of these papers were examined. For some papers, just reading the title and abstract was not enough to decide whether to keep them in the selected papers' pool.

We kept those papers for the next round. A total of 1617 papers were discarded based on the inclusion criteria described in Table 2.

We read the title, abstract and keywords of each paper in the Journal of Computer Security, the ACM SIGSAC series of Conferences on Computer and Communications Security, and the RSA series conferences and filtered 19 papers. After round 1, we selected 290 papers. Then we removed the duplicates and excluded the papers that were shorter than 6 pages. Finally, we screened the whole text and applied the eligibility criteria to select the relevant papers. A total of 37 papers were selected from the digital libraries. To ensure the inclusion of the relevant papers that we might have missed, we extended the search in Google Scholar. We searched for the string "Security and Orchestration" and checked both the titles and abstracts of the top 200 results. We only included 6 articles that were not found in the automatic and manual search procedures of phase 1. We applied all the eligibility criteria while selecting the papers from Google Scholar.

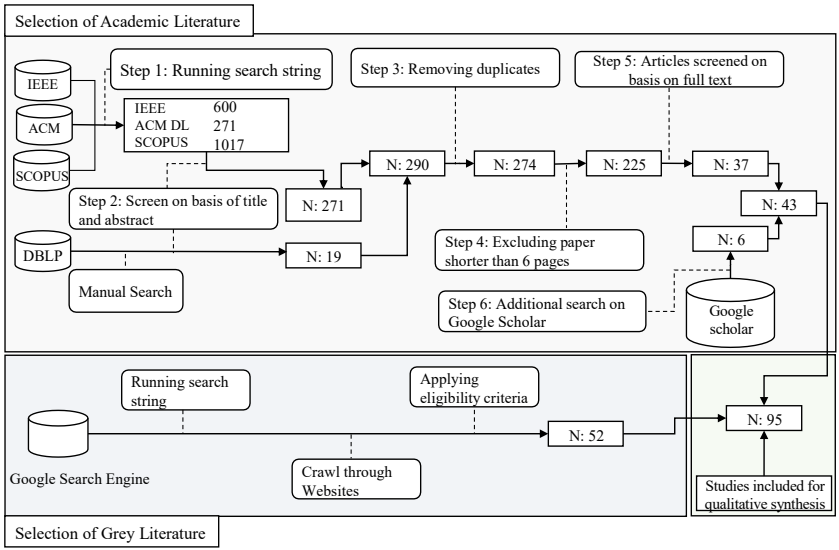

**Fig. 3** Study selection process of our MLR

*2.4.2.Selection of Grey Literature.* In the next phase, we used the Google search engine and checked the first 10 pages. We only continued further if needed. We identified several practitioners (niche and start-up) who were contributing to the field of security orchestration. We crawled through their websites and looked for the relevant resources and white papers. We applied the eligibility criteria while selecting the papers. At the end of this process, we identified a total of 52 papers that include white papers, blogs, news articles, and websites.

Finally, we included 95 pieces of literature (Fig. 3) for data extraction and synthesis. Fig. 4 showed the distribution of the selected pieces over several types of venues. For both cases, we excluded the papers published before 2007. Table 3 enlists the pieces of work that was finally reviewed.

**Table 3.** Study selected for data extraction and qualitative analysis

| Academic Literature | Grey Literature | |
|---|---|---|
| | Websites & Blogs | Whitepaper |
| [16, 17, 30-33, 42-78] = 43 | [11, 13-15, 18, 20-23, 79-100] = 31 | [12, 19, 29, 101-118] = 21 |

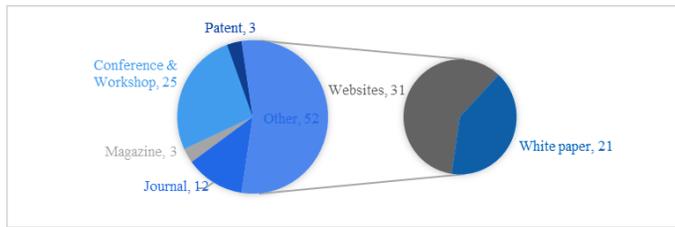

**Fig. 4** Distribution of selected articles over venues

## 2.5. Data Extractions, Synthesis, and Analysis

Following the process of MLR, at this step, we read, assessed, extracted data, and summarized the findings from the selected papers based on the pre-defined RQs and motivators (Table 1).

*2.5.1. Data Extraction.* We identified and extracted the relevant data using a pre-defined data extraction form from each of the selected sources that we needed to answer the research questions. We also extracted some general information, e.g., authors name, venue published and published year. We conducted a pilot study on a set of 10 sources before deciding about how to extract the required data. We stored all the extracted data in a spreadsheet for analysis.

*2.5.2. Synthesis and Analysis.* The extracted data were stored in different sections of the data extraction form (a) security orchestration definitions, functionalities and application, (b) challenges to be solved and (c) security orchestration practices, tools, and techniques, to perform the synthesis and analysis of the extracted data. We analyzed each set of the data items using qualitative analysis methods. We used a combination of different qualitative analysis methods (i.e., narrative synthesis and thematic analysis). For example, for classification and categorization of data, we used thematic analysis [119, 120]. We followed several steps to analyze the data including getting familiar with the extracted data by carefully reading each piece of the data. We collaboratively analyzed and systematically synthesis the extracted data for developing a taxonomy to report the results in a generalized form. The taxonomy developed in this study has been used for reporting the functionalities, benefits, and aspects of security orchestration in this paper. For data analysis, we followed the qualitative data analysis guidelines [120]. We have included a table of abbreviations, Table 6 at the appendix, used in this paper. We report the synthesis result in section 3, 4, 5, and 6.

## 3. Security Orchestration: Definitions, Functionalities, and Elements

This section presents the findings for RQ1 *"What is Security Orchestration?"*. Our data analysis for RQ1 reveals some key definitions of "security orchestration" given by practitioners, the functional and non-functional requirements, and the key functional components of security orchestration.

### 3.1. Definitions

Our analysis shows that practitioners widely use the term security orchestration with no clear and common definition. We assert that having a common working definition of security orchestration will help practitioners and researchers to define a discipline of research and practice for promoting practices, processes, and tools. The term security orchestration is being mostly used as a Buzz-word that can lead to misinterpretation of the core concept of orchestration [18, 83, 84, 102]. Some organizations and practitioners confuse security orchestration with security automation [23]. We present a few key definitions from the reviewed work.

According to HEXADITE, "Orchestration is the practice of connecting existing security tools together through APIs to streamline incident response processes." Here Hexadite has considered orchestration more as an integration tool and presented a definition for security automation. Barak Klinghofer, CPO of HEXADITE [13] has defined "The active process of Mimicking ideal steps of a human would take to investigate a cyber threat, determining whether the threat requires actions, performing necessary remediation actions, deciding what additional investigation should be next" as security automation [104]. According to a start-up company KOMAND [11], security orchestration is more than just connecting security tools. Their definition [11] is -

> *"Security orchestration is a method of connecting security tools and integrating disparate security systems. It is the connected layer that streamlines security processes and powers security automation*."

Markets&Markets [91] state that "Security orchestration is an approach to automatically respond to security incident and protect IT systems in organization from advanced cyber-attacks and vulnerabilities". Microsoft has distinguished security automation and security orchestration [21] -

> *"Security Automation is the use of information technology in place of manual processes for cyber incident response and security event management"* and *"Security orchestration is the integration of security and information technology tools designed to streamline process and drive security automation".*

ThreatConnect [86] has presented distinct definitions for security automation and security orchestration-

> *"Security automation is the automatic handling of a task in a machine-based security application that otherwise be done manually by a cybersecurity professional"* and *"Security orchestration is the connecting and integration of various security application and process together" [86].*

ThreatConnect has defined security automation and orchestration [86], "Security automation and orchestration is a coordination of automated security tasks across connected security application and process."

According to Forrester, security automation and orchestration should be described together as technology products. They have defined security automation and orchestration [83] as-

> *"Technology products that provide automated, coordinated, and policy-based action of security processes across multiple technologies, making security operations faster, less error-prone, and more efficient."*

Clearly, Forrester's definition asserts that automation can help take the full benefits from security orchestration. Bruce Schneier [14], chief security officer of IBM, has stated security orchestration as the unification of people, process, and technology. He claims that security orchestration is keeping people in the loop of security automation where the computer performs the automation of certain activities, but a human coordinate the activities. It is more about making people effective. He also pointed out that the security incident response needs to be dynamic and agile. DFLabs's Oliver Rochford has defined security orchestration as the junction where people, process, and technology all come together [18]. According to him, people build automation into the process and consume information and insight generated by technology. Security orchestration is the realization of three paradigms – integration, orchestration, and automation. Our definition of security orchestration is:

> *"Security Orchestration is the planning, integration, cooperation, and coordination of the activities of security tools and experts to produce and automate required actions in response to any security incident across multiple technology paradigms."*

This definition provides siloed security solutions the ability to share information and threat intelligence among them through a unified platform. This is achieved by seamless monitoring, situation awareness, data analytics, knowledge representations, and semantic knowledge sharing among the existing security solutions. It is clear that there is a need for extensive training to learn from human behavior to provide AI capabilities that can enable an enterprise for long-lasting development and deployment of security solutions using the existing tools and protocols. Security orchestration works as an intelligence assistant for security experts.

## 3.2. Functionalities of Security Orchestration

In this section, we report the findings of the functionalities of security orchestration. Several reports (i.e., [83, 101]) have mentioned security orchestration as one of the emerging technologies, which has the potential of being widely adopted in near the future [101]. One of the motivators of the security orchestration is to bridge the gap between detection and remediation of security incidents [13, 111]. Most of the detection solutions are automated where the response processes are still reliant on human. To bridge this gap, there is a need to unify the activities of security tools, streamline the workflows, and choose the right course of actions. According to Demisto [115], a comprehensive security orchestration platform must be able to automate security tools activities, create playbook with complicated logic, and track and orchestrate the tasks assigned to an analyst. Paul Weeden [94] has stated: "*Security orchestration makes the most of human skills by bringing together automated tools and reports to provide risk information exactly when and where it is needed.*" Fig. 5 highlights the key functionalities of security orchestration in three paradigms, integration where a security orchestration acts as a middleware, orchestration that is the process of translating complex process into streamlining workflow and automation that enable an automated response.

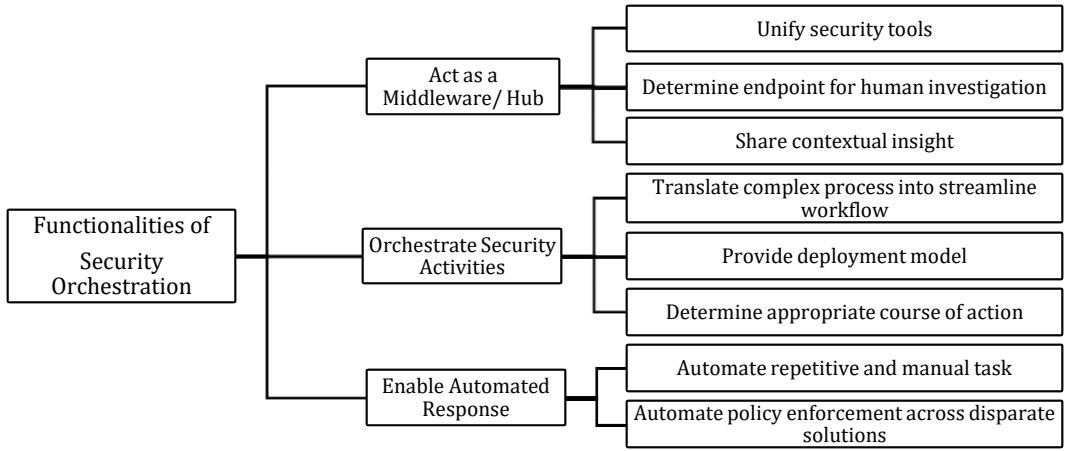

**Fig. 5 Key functionalities provided by Security Orchestration**

*3.2.1.Middleware/ Hub.* Vendors have mentioned security orchestration as a platform that acts as a hub for unification, coordination, data sharing and analysis for disparate cybersecurity and IT technologies [12, 16, 106]. Security analyst can easily integrate multivendor security tools, share threat intelligence, collaborate with the external organizations to get an insight of an organization's security state through an orchestration platform. SDSec orchestration solution has a layer of functionalities to perform communication and coordination among different subsystems [31].

**Unify security tools:** Several of the reviewed work mentioned that security orchestration unifies disparate security tools and processes [11-13, 23, 30, 62, 108, 121], integrates enterprise security architecture [101, 115], connects detection, networks and endpoint security systems [17, 111], and perform coordination among security tools activities [16, 19, 23, 43]. Connecting the activities of disparate security tools makes the incident handling process efficient and effective for security analyst. Security orchestration platform provides a single console or platform to integrate security tools activities, removes the operational silos, and helps security experts to free their time [93, 109]. Unifying intelligence according to vulnerability also minimizes the overall complexity of the incident response process [30, 43, 53]. Through the coordination platform, security tools can interoperate with each other to enhance organizational protection and defense systems [62, 95, 112]. Feitosa et al. [49] propose collaborative solutions to detect intrusion and anomalies by analyzing the co-creation of events and alerts among different subnetwork. The orchestration tools proposed in [16] are designed to coordinate the safeguarding function by calling an individual software package with respect to an installation. Jeong et al. have proposed a coordination module for organization architecture integrating cyber forensic functions [78].

**Determine endpoint for human investigation**: A security orchestration platform [83] can enable security experts to get insights into several security controls activities, operate disparate tools as a unified system [17, 19], and to collaborate with other experts for planning and decision making [11, 93, 95]. The security orchestration solution informs and educates security analyst about threat behaviors [83] and notifies about the supported policies [111]. By orchestrating various activities, a system can decide when human insight is required [11, 13, 75, 84, 102, 110]. For example, the work in [42] also highlights the critical assets with high priority to the administrator for investigation. The motive is to keep analyst focus on threats that demand their immediate attention and expertise.

**Share contextual insight:** A single security prevention and detection system usually suffers from the tunnel vision syndrome that leads to an inability to detect certain types of attacks such as Distributed Daniel of Service (DDoS). A security orchestration platform gathers threat intelligence from various external sources (e.g., web pages and blogs), extracts key features from a huge volume of threat intelligence data, and provides the contextual insight related to alerts or attacks to a security analyst. In addition, it engages security tools to perform complete monitoring of endpoint [54, 101, 112], correlates their activities, and provides real-time visibility of known and unknown threats to security analyst [93, 96, 108]. An organization can share contextual device data with the third-party system [12, 111]. It helps security analyst to reduce and mitigate the risk exposure [108, 111], make a faster decision based on context [17, 75, 81, 110], and gather an overview of what is happening in various subnetworks within an organization [12, 19, 20, 101]. By sharing the contextual insight, an orchestration platform works as a collaborative platform that also enables training the analyst based on past investigations [19, 81, 93]. The online evaluation framework proposed in [42] provides situational awareness to an organization so that it can take appropriate actions. By assessing the security state of an organizations' different assets, the proposed framework helps administrator to identify compromised assets and prioritizes alerts [42]. A set of papers [30, 46, 48, 55] has proposed a platform for security experts and security solutions used to share their knowledge. Jeong et al. [78] have followed the structure of having a coordination group with a participant group to propagate the relevant information to the external work or another coordination group. RiskVision has proposed security orchestration solutions to unify stakeholders in business, IT and security solutions to provide automation for end-to-end cyber risk prevention and response [95].

*3.2.2. Orchestrate Security Activities.*

***Translate complex process into streamline workflow***: After receiving alerts, security experts need to perform multiple steps to find the attacks, vulnerabilities, affected endpoints, and mitigation solutions. These steps include the complex process of data collection, investigation, remediation, evaluation of actions, and deciding the appropriate course of actions. Several papers [11, 13, 23, 29, 42, 86, 104, 106, 115] have mentioned that the motivation of orchestration is to translate the complex process of threat investigation into a streamlined workflow through automation and orchestration. A streamlined workflow requires a standardized process that includes proper planning for incident response, policy execution, investigation, response action, and remediation process [31, 105, 108, 109]. The workflow is designed to mimic human activities of threat investigation to reduce the cumbersome manual process, human errors, and improve staff capabilities to incident response. Orchestrating and integrating security tools' activities allow experts to simplify complex workflow, coordinate the flow of data and tasks, and enable the powerful machine to machine automation [11, 102, 108, 118, 121]. The task can be fully or partially automated based on the complexity of the threats [109]. ForeScout has proposed a rule engine and a workflow engine to make instant decisions and offered data aggregation to provide in-depth awareness about the environment [101]. The online Evaluation framework, Seclius [42], translates alerts into system security measures to reduce the reliance on human expertise on capturing system characteristics through low-level alerts. This work also provides a raking of the affected systems assets and malicious events for organizations to help security administrator [42]. Similar to this, the Premise-aware Security Instrumentation (PSI) policy engine proposed in [33], translates the high-level security postures provided by an administrator into per device intents.

***Provide deployment model***: Several security vendors provide orchestration deployment services that require appropriate orchestration and automation of existing security tools along with organization external and internal infrastructures [47, 51, 80, 85, 89]. FireEye deployment service for automation provides functionality to manage events across multiple FireEye and third-party products and ensures deployment is successful [80]. Deployment model depends on organizations' scale, complexity, and course of actions. Vendors are providing flexible deployment model for organizations to ensure simple installation and management of various infrastructure. This action also ensures efficient deployment in heterogeneous environments. Organizations can choose security policies based on their need to restrict the access and tailor security configurations [29, 113]. Additionally, the testing and evaluation of deployment model are also done once any change has been made. The proposed system can provide a progressive deployment module to perform upstream rules filtering that helps to reach the source of attacks [46]. The work in [47] proposes an innovative solution to perform quick deployment of various security mechanisms. The orchestration system proposed by [51] arranges appropriate virtual instances in the right place - virtual appliances are automatically added and controlled. It also automatically moves traffic to virtual network to prevent major harm, blocks the attack, and strengthens the system.

***Determine appropriate course of actions***: A security orchestration platform can help promptly resolve an incident to determine the appropriate and effective course of actions [13, 31, 80, 85, 92, 93, 115]. By choosing the appropriate course of actions, security orchestration maintains process consistency across a security program. FireEye has mentioned the in-built course of actions with automated support for all the needed steps for handling a security incident as a core of an orchestration platform [80]. Also, various kind of alerts (i.e., phishing, and endpoint contamination) are needed to distinguish remediation activities with different courses of actions. Upon investigating an alert, an orchestration platform can determine the proactive response to threats, or may initiate an additional investigation based on an attack's complexity [80, 84, 111]. In many cases, a post-attack investigation or evaluation task can also be instantiated. Feitosa et al. [49] have proposed a framework, *"Orchestration-oriented Anomaly Detection System (OADS)"*, that performs coordination and collaboration among different anomaly detection techniques to detect and evaluate threats and choose right actions. Security experts do multiple investigations in response to an alert. In the process of orchestration, one investigation usually triggers multiple investigations [13].

Workflows are designed to choose the appropriate course of actions, simplify threat response through integration and automation [11, 12, 108], perform necessary remediation [111], decide additional investigation [104], design documents for playbook review [80], and define sources of information to help expert in solving the identified problems [107]. A formal workflow helps maintain effective communication and strong collaboration among cybersecurity teams [104]. Providing a formal workflow helps security experts and orchestration platform to maintain consistency across actions [108]. This simplifies and accelerates alerts investigation, eases proactive hunting of attackers, accelerates Return on Investment (ROI), and eliminates the need for continuous assessment. The security functions are dynamically inserted into the workgroup based on the policies [29]. The operating principle of the framework in [17] has security controls that operate on their own and perform functions of context sensing, policy decision and policy enforcement. Each function is logically independent of each control. The orchestration tool proposed in [16] works as an interface to perform security scanning and testing or other security functions. This tool enables different safeguard software packages to come to an agreement for invoking the necessary function(s) owned by any of the software packages. The framework discussed in [78] helps to effectively implement the coordination of an organization's functions by performing an on-site and online investigation to provide security warnings and appropriate response actions.

*3.2.3. Enable Automated Response.* Security orchestration automate incident response activities. HEXADITE has automated 800,000 man-hours of work in two years that is equivalent to *$38.5* million in customer savings [13]. Several papers [12, 13, 81, 101] have reported that security orchestration automates the entire threat defense life cycle and provides intelligence automation services. ETSI (European Telecommunication Standard Institute) has considered automating the control of deployment and configuration of the security functions as a substantial prerequisite of orchestration [52]. According to Forrester, orchestrating the incident response activities enable automated response without the need for coding skills [83]. Orchestration allows autonomous control and protection of network through discovered insights [111].

**_Automate repetitive manual task_**: Vendors use a security orchestration platform to automate repeatable tasks and remove duplicate incidents to optimize security staff's capability and reduce the overall cost [11, 21, 81, 90, 96, 108, 111, 113, 121]. Automating the routine tasks help security experts to tackle more critical problems [11, 12, 23, 86, 104, 111]. According to Swimlane, 80% to 90 %of all security operations of an incident response can be automated to some extent [117]. The collaborative incident response planning process design discussed in [45] helps practitioners to come up with the repeatable and executable planning process. Some vendors provided a platform that also automates the deployment of security functions through a network infrastructure [29]. Several papers [30, 47, 84, 95] have proposed to automate the analysis of cyber threat intelligence, that includes extracting data from technical blogs, websites, finding correlation among different reported attacks, and updating incident severity based on threat intelligence feeds. An orchestration platform decreases the response time by minimizing error-prone manual process and codifying real-world expertise. Koyama et al. have also reported a security operations automation framework that helps in optimizing decisions with regards to a variety of security sensors and appliances [30]. Luo et al. [31] have automated cybersecurity operations in the software-defined environment.

**_Automate policy enforcement_**: Security policy enforcement greatly benefit from automation that considers all the tools, devices, and measures required for a security policy implementation. Security orchestration enables an organization to automate policy enforcement and configuration at runtime [29, 32, 33, 52, 60, 62, 106]. A set of systems developed to derive policy decision based on contextual data and provide real-time policy enforcement, have been reported in [17, 32, 60]. Yu et al. [33] have introduced a multistage mapping mechanism to automatically adjust policies based on network devices. In [32], policy enforcement is performed automatically. ETSI has aligned the security policies in an automated way inside virtual, physical, and hybrid network [52]. Dynamic enforcement of policies allows automatic configuration of security elements and update of threat intelligence [29]. An organization can automate policy enforcement across disparate solutions [111]. Luo et al. [31] have proposed to provide consistent security policies by orchestrating software defined security services across a heterogeneous cloud environment. A generic security orchestration framework proposed in [17] enables ad-hoc, context-aware policy criteria to be applied in real time by using an ecosystem of security control connected via Data Exchange Layer (DXL).

## 3.3. Quality Requirement for Security Orchestration Platform

We have identified the key quality requirements of a security orchestration platform. Fig. 6 shows the main quality attributes of security orchestration platform gathered from the existing literature.

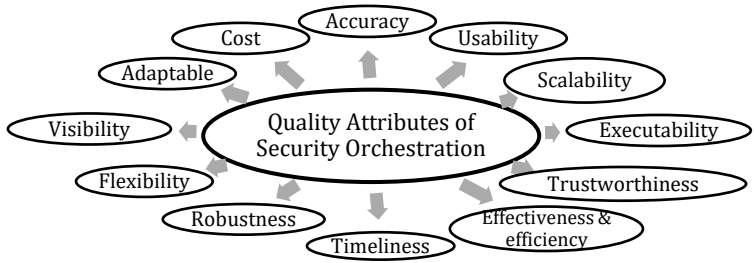

**Fig. 6 Quality Attributes of Security Orchestration Platform**

These are the functional and non-functional requirements of security orchestration system. Every organization needs to consider these attributes before adopting or implementing a security orchestration platform. The quality requirements of a large-scale system are expected to guide the key architecture design decisions. Hexadite has mentioned the pre-requisite for security orchestration which is basically the quality requirement for security orchestration platform [105].

Table 4 enlists the quality attributes and a set of corresponding metrics that can be used to measure these attributes. For example, usability is a crucial factor for the effective utilization of a security orchestration system. A security orchestration platform requires to have a simple and powerful user interface that can be easily customized for different types of security orchestration users. An orchestration platform should have a flexible architecture so that each user can create a work environment according to their need for a service.

## 4. Key Components of Security Orchestration

Organizations and vendors must consider the key components of security orchestration platforms before adopting them. We have identified several core components of a security orchestration platform. Most of the reviewed studies have a combination of these components that we have categorized in three classes – unification, orchestration, and automation unit. This classification is based on the functionalities discussed in *section 3.2*. We have considered the external security tools as another key component of an orchestration platform because most of the orchestration vendors consider that an organization already own some security tools and use the functionalities of the existing security solutions. Fig. 7 presents the details of the classification of the security orchestration core components. All the modules proposed here comprise of learning capabilities, specific policies, and storage to store the threat data. These modules also store security policies and rules associated with various organizational assets and endpoints [31]. We consider organizations' devices that can be a server, client personal devices – laptop, mobile, personal computer, or organization owned workstation and so on as endpoint. It helps a system to gather knowledge about the new threats and threats patterns. We do not include these in the presented classification to keep it simple.

**Table 4 Quality attributes of a security orchestration**

| Quality Attribute | Measurement Metrics | Articles |
|---|---|---|
| Accuracy | ▪ Accuracy of diagnosis<br>▪ Appropriate measure against attacks<br>▪ Accurate classification and reliable taxonomies of threats<br>▪ Data integrity | [17, 30, 42, 43, 47, 49, 50, 56, 59, 62, 63, 66, 67, 74, 95, 101] |
| Usability | ▪ Ease-of-use, easy to manage, connect, & repeat task, interruptible<br>▪ Simplified user interface to control security tools<br>▪ Simplification of security management task for network administrator & auditing module<br>▪ User satisfaction<br>▪ Higher analyst productivity<br>▪ Accessible and stable threat intelligence | [13, 16, 19, 31, 42, 44, 45, 51, 53, 65, 73, 75, 81, 86, 102, 105, 110] |
| Scalability | ▪ Vendor agnostics,<br>▪ Independent security policy orchestration<br>▪ Extensible architecture | [13, 29, 31, 32, 42, 44, 46, 48, 63, 65, 73, 84, 89, 101, 110] |
| Executability | ▪ Qualitative and quantitative information about the current incident<br>▪ Measurable security system,<br>▪ Measurable goal<br>▪ Security state of different assets of organization infrastructure | [11-13, 22, 45, 53, 74, 110] |
| Trustworthiness (Reliability) | ▪ On human: expertise level, fairness to collaborator, reputation<br>▪ On existing security tools – trust value, Predictability | [17, 30, 42, 43, 45, 47-50, 53, 56, 59, 62, 63, 66, 78] |
| Effectiveness & Efficiency | ▪ Increase in detection rate,<br>▪ less overhead, do more work with existing staff<br>▪ Reliance on human ability & satisfaction, optimize resource & performance<br>▪ Predictable cost structure, indicator of compromise;<br>▪ Key indicator to measure security effectiveness: Mean time to notification, remediation, & investigation<br>▪ Speed of integration and speed of deployment | [12, 17, 19, 29, 42, 45, 54, 58, 59, 62, 65, 67, 74, 77, 81, 92, 101, 105, 110, 117] |
| Timeliness/ Speed | ▪ Time to perform raid recovery<br>▪ Time to detect, triage attack and remediation<br>▪ Time need to analyze an attack,<br>▪ Time for policy enforcement,<br>▪ Delay in business activities,<br>▪ Overall latency of packet processing, | [12, 13, 17, 19, 20, 28, 30, 44-46, 49, 51, 62, 65-67, 73, 75, 84, 86, 90, 101, 105, 110] |
| Robustness | ▪ Robustness to DDoS<br>▪ Capacity of attack detection,<br>▪ Incident response capacity | [12, 32, 46, 48, 61, 62, 110] |
| Flexibility | ▪ Feasible to update<br>▪ Flexibility to design workflow automation<br>▪ Flexibility to adapt process & accelerate response to all type of threats | [17, 20, 31, 32, 49, 51, 63, 65, 81, 84, 90, 96, 101, 110] |
| Visibility | ▪ What analysis are available and what are their abilities<br>▪ Security state of the organization<br>▪ Secure configuration guidelines | [12, 13, 17, 29, 31, 50, 53, 55, 64, 81, 90, 95, 110] |
| Adaptable | ▪ Compatibility with existing network topology and security appliance<br>▪ Adaptable with current process | [11-13, 33, 52, 56, 77, 90, 110] |
| Cost | ▪ Low computation cost<br>▪ Cost effective security orchestration platform for mixed environment<br>▪ Additional resource<br>▪ Cost of ownership | [12, 13, 29, 33, 44, 49, 86] |

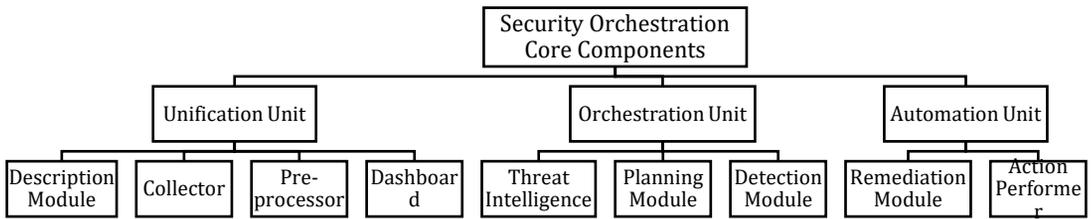

**Fig. 7** Categorization of core components of security orchestration

### 4.1. Unification Unit

Under this category, we have considered the components that are designed to unify the existing security tools activities. The unification unit works as middleware or hub as briefly discussed in 3.2.1.We consider data collector, alert pre-processor, and description module under this category.

*4.1.1. Description Module.* This component refers to language and models to represent configuration, deployment, and control tasks of security orchestration. Luo et al. [31] proposed a system that requires an abstract service to be defined for the same type of security services. Security management is built using the API of abstract service. Luo et al. [31] proposed security control capability descriptor, that describes all types of security tools, inputs, and necessary attributes. The description module requires well-designed API to connect the existing security tools. The security service requirements and descriptors are derived from organizational security policies and controls respectively. The set of interfaces are mentioned as connectors in [31]. They have introduced two sets of interfaces: event connector and command connector. Both security and network events are received by the event connector [31]. These events come from external sources. The events connector also sends the information to the playbooks. The set of interfaces under command connector sends a command to a security control to modify their configurations and update operation behavior [31].

The orchestration platform needs a suitable API to help third-party integration and control the activities in different layers [63-65, 80, 93, 97]. Intel has proposed a bi-directional notification based API to orchestrate virtual security in Software Defined Data Centre (SDDC) [29]. The global threat intelligence platform supports dedicated tools to provide a simplified interface to a firewall's control [30]. Swimlane also uses API to enable one-click automation [19]. As stated by Bernd [52], one of the main purposes of ETSI's management and orchestration group is to control Network Function Virtualization (NFV) environment through virtualization and automation as much as possible. This work has enhanced NFV reference architecture with security orchestrator, the interaction of the security orchestrator with the existing NFV orchestrator and Virtual network function manager of the reference architecture. The workgroup also included tasks of the security orchestrator and the required interfaces to interact with. In this work, the author has stated how security is managed through orchestration in a virtualized network environment. The correlation module of alert correlation architecture proposed in [43] uses application interface with the reasoner for reply and request. The DXL layer proposed by McAfee for enterprise service bus works as a connector for diverse security elements and has an extensible data exchange framework to facilitate configuration of trustworthy data representation [17]. Safeguard interface module of enterprise-level security orchestrator provides a layer of software for a consistent interface to abstract away the changing nature of the underlying safeguard software packages.

*4.1.2. Collector.* Most of the reviewed security orchestration platforms have collectors to collect all the necessary information for integrating security tools or devices to its system. In several studies, network traffic and alerts are collected and pre-processed before analyzing and taking a decision [21, 44, 49, 60, 66, 90, 94, 114]. The collector collects both raw context and structured format data [64]. The orchestration server engine of the security orchestration framework presented in [17] works as a collector and receives contextual data from clients. The orchestration platform discussed in [49] has OADS miner as a core component. The OADS miner works like a consultant to the overall system, which comprises of OADS crawler. The OADS crawler is designed to gather new information from the Internet about threats, vulnerabilities, attacks, the origin of attacks and store them in a unique repository. The threat intelligence unit works as a blog scrapper that crawls through technical blogs to collect, gather, and share threat intelligence data [47, 64]. Security orchestration utilizes global threat intelligence platform to collect external threat intelligence and to prevent data infiltration (the action of entering or gaining action) and subsequent actions performed by attackers [30].

*4.1.3.Pre-processor.* A pre-processor receives raw alerts from several security tools and prepares the alerts for analysis. The alert pre-processor first decides the alert adequacy, and then it aggregates the alerts into clusters based on similarity. A pre-processing of threat data involves sentence splitting, special content extraction, content term location, topic classification, template removing, and content sanitizing [47, 60, 63, 64, 74]. Feitosa et al. [49] have used the Intrusion Detection Message Format (IDMF) standard to aggregate the alerts of several IDS tools. The aggregation also helps to explore the distance between the times of different alerts, determine the alerts field, and make a hypothesis about alerts and defense strategies. The proposed architecture for collaborative Intrusion Detection System [44] also uses IDMEF to unify alerts from multiple IDS. In [31], the orchestration platform has a modular physical logical attribute mapping that maps all assets' physical attributes to its corresponding logical attributes. The study in [43] has combined several knowledge representation languages, for example, IDMEF, TAXII, OVAL, STIX, NVD, to propose ontological conceptualization and divide the knowledge into several groups. Their proposed ontology-based event correlation architecture consists of two essential modules, one is conversion, and another is the correlation module. The conversion module consists of parting reports, translator and ontology.

*4.1.4.Dashboard.* The dashboard category consists of tools aimed at visualizing the activities of an orchestration platform. According to Demisto Inc. [114], the dashboard will bridge the gap between security operation center and technology used to keep the organization secure. The dashboard provides an aggregated view of different scenarios, assets, and metrics [85, 95, 114]. FireEye uses a centralized dashboard to facilitate advanced threat hunting [80]. In Enterprise Level Security Orchestrator [16], the security orchestration tool provides administration an interface through a dashboard. The dashboard can be designed to provide an integrated view of an organization's overall system to help experts understanding the security states; for example, security experts can see all the scans in progress from a single console [16, 53, 75]. FireEye orchestration deployment service provides documents associated with cyber playbook review that helps orchestration operation team to understand the playbook [80]. Most of the reviewed systems generate some form of alerts reports for security experts [44, 65, 74]. Through these reports, security experts can get a high-level overview of an organizational security system. These reports also enable experts to identify the security state of critical assets and the affected networks or subnetworks within an organization. The reporting tools receive a recommendation from the remediation engine related to threats. Threat visualization and analysis is an important part of security orchestration. A set of papers have mentioned several web portal or public websites that provide a web interface to visualize the threats [30, 55].

## 4.2. Orchestration Unit

One of the substantial pre-requisites of a security orchestrator is to automate the control of deployment and configuration of all the security requirements. For this category, we consider all the components that are required to perform the functionalities described in *section 3.2.2*. For example, the security orchestration framework describes in [17] has provided a security orchestration engine to receive contextual data from clients. It comprises one or more logic element(s) which are designed to work with the contextual data. Koyama et al. [30] have proposed a three steps process to cope with new sophisticated unpredictable threats – collect, judge, and control. The operating principle of the proposed framework in [17] has security controls, which perform context sensing and based on the context generate and enforce security policy decisions. A security orchestrator needs to manage several activities as stated in [52, 58]. The proposed security orchestration solution performs central management of security service and trust. FireEye in their orchestration platform has used a specialized component, called case management for managing various cases. We categorize the orchestration unit into three modules threat intelligence, planning module and detection module.

*4.2.1.Threat Intelligence Unit.* Cyber threat intelligence can be considered as a database of evidence of existing and emerging attacks [47, 53, 64, 65, 73]. Threat intelligence consists of information related to attacks context, adversary strategies, mechanisms, indicators of compromise, possible course of actions, tactics, and techniques [47, 85, 86, 91]. Threat intelligence plays a key role in security orchestration. An organization can gain the visibility of threat landscapes by using threat intelligence. It helps organizations to identify the early sign of attacks [47]. Organizations are collecting and exchanging threat intelligence data across several domains and stakeholders as Indicator of Compromise (IOC) [47, 61, 62]. Example of IOC can be forensics artifacts, virus signatures, IPs/ domain of botnets, and MD5 hashes of attacks files. Most of the security orchestration platforms have considered threat intelligence as an essential element to identify attack behavior at an early stage [30, 49, 51]. Security orchestration with Global Threat Intelligence Platform (GTIP) [30] incorporates proactive defense technologies including threat intelligence. There are several open source threat intelligence platforms that provide high accuracy and coverage [47]. Each has their own specialized techniques for data extraction. A threat intelligence data may also suffer from quality issues such as accuracy, completeness, consistency, timeliness, and relevance [53]. Filtering, configuration, and searching options are not available in some of the current threat intelligence tools. Xiaojing et al. have identified 45 blogs that are operated by renowned organizations and practitioners to cover major security incidents [47]. These blogs consistently publish verified IOCs that a security orchestration platform utilizes for updating the threat intelligence information.

*4.2.2. Planning Module.* We consider the cybersecurity playbook as a planning module that outlines the steps to respond to an incident, including incident qualification, triage, investigation, containment, notification, and post-attack analysis [23, 84, 86, 94, 96]. A playbook arranges security operation into a human-led security workflow that is a coordinated set of activities performed by various components to complete an incident response within an organization. According to an organization's policy and infrastructure, a security playbook creates smart branching workflows and also supports the activities of SIEM, firewall, threat intelligence, IPS, vulnerability reporting, and ticketing systems [66, 73, 114]. Incident response playbook contains various course of actions. The FireEye has proposed incident response playbook as one of the key features of security orchestration [80]. To provide continuous proactive security, FireEye designs appropriate orchestration platform according to an organization's requirements, deploys and tests it in the organization's environment, integrates security tools with third-party solutions and operates it to execute appropriate plan against a security threat. Orchestration of Software-defined Security Services [31] design playbooks to store the actions (operation plans) related to important security events (security alerts). The security administrators take necessary steps based on the actions mentioned in the playbooks [31].

Zonouz et al. in [42] have proposed a consequence tree (a tree of critical assets defined an administrator) to capture critical IT assets and organization security requirements. Each organization has their own list and priority of critical assets. The consequence tree is built using this list of critical assets. Kamal et al. [45] have considered the Incident Response Plan (IRP) as a crucial component of collaboration engineering. The authors highlighted that creating IRP through collaboration among a group of experts is challenging when the time is very short. The PSI policy abstraction helps an administrator to define policies in terms of what they do rather than the details how to do them [33]. With the help of PSI engine, an administrator can define how the traffic of a particular device should be processed and where to forward them. The security orchestration platform requires a proper planning of the incidents. Without proper planning and preparation, a security orchestration platform ends up automating a lousy process that might slow people down [23].

*4.2.3. Detection Module.* The detection module detects the anomalies and attacks around the organizations based on the gather and pre-processed data, shared insight, and knowledge of the playbook. Analyzer and the decision service unit are the two-main core components of the detection module. In the following paragraphs, we have briefly described these two components.

***Analyser***: The analyzer receives aggregated alters from the alerts pre-processors, correlates them, validates the assumptions and if possible, predicts future threats and targets. The analyzer performs an automated analysis of a system, analysis of logic unit, and enrich data to make sense of it [28, 60, 62-64, 84, 86, 94]. A set of papers have reported correlating suspicious evidence provided by distributed security entities to identify distributed attacks [43, 44, 46, 47, 66, 74, 84]. Enterprise Level Security Orchestrator [16] can analyze alert data and detect threats. It stores all the threats data in its analysis storage. It generates a set of rules or traffic patterns as a decision table by finding the correlation among different alerts. Similarly, the PSI performs packet pre-processing and event pre-processing before analyzing the data [33]. Kenaza et al. [43] have used ontological reasoning approach to correlate alerts. The proposed ontology-based event correlation architecture has a correlation module, that works as a reasoner. Feitosa et al. [49] proposed a collaborative solution to detect intrusion and anomalies by analyzing the co-creation of events and alerts among different subnetworks. It derives policy decision based on the contextual data it receives from an orchestration engine [49]. The solution proposed by [47] tried to find correlation among IOC data to check relation among threat data. A core part of their solution is the Analyzer. Seclius [42] tracks the interaction among files and process to probabilistically identify dependencies among assets of an organization.

***Decision Service Unit:*** Most of the reviewed security orchestration systems have decision services unit that orchestrates the activities for automated decision-making [78, 81, 84, 95]. The decision service unit makes security policy decision(s) related to vulnerability and threat assessment and assessment of security enforcement system [60]. The decision service unit receives summarized information from analyzer and collector about suspicious behavior and generates decision-based on them [49]. The finite state machine (finite automata, Markov chains or stochastic regular grammars) is a popular method used in the decision process. For example, the security orchestration framework in [17] uses the policy orchestration state machine to provide policy decision to the security orchestration state machines and deriving the policy decision based on the contextual data received from the security orchestration server engine. Policy decision logics are extracted from individual control. The decision logics enable additional, ad-hoc, smart logic, and intelligence analytics to be injected into the real-time policy decisions. Thus, policy decision logics capture context and drive actions staged over multiple points in space and time [17]. Similarly, Seclius [42] has constructed dependency graph and consequence tree of existing assets to probabilistically determine the comprised assets, prioritize alerts, and provide security state of different assets to the administrator.

Koyama et al. [30] have used an optimal decision-making technology and diverse threat intelligence with a variety of security sensors and appliances to choose the correct countermeasure for stopping an attacker's internet-based actions. Utilising the workflow engine is also a popular strategy to take a decision. For example, Rochford et al. [28] have decided the actions based on the workflow engine. The reviewed study, SOSDSec, generates a security service binding upon finding matching among security requirements and abstract services [31] which contains the information related to assessment and its service provider. A change in the security control and module causes a change in the security service binding. Similarly, the differentiated search engine in OADS miner discussed in [49] is used to generate the decision based on the queries received from end users or system tools. The decision service unit makes all the decisions related to the OADS miner like activation, deactivation, and parameter change and stores the configuration parameter in a file. The decision service unit also provides recommendation after analyzing information about attacks. The alert buffer of security orchestrator proposed in [16], continuously sends updates to a dashboard.

### 4.3. Automation Unit

The automated unit performs all the automated task based on the decision generated by the decision unit and analysis of the workflow. The remediation unit and actioners performer help a security orchestration platform to deal with the automated task. In the following paragraphs, the role of the remediation module and actioners performers have been described in detail.

*4.3.1.Remediation Module.* The remediation module promptly configures countermeasure and security operation based on the decisions of the detection module to remediate threats [32, 65, 84, 90, 94]. A remediation module reported in [30] performs automatic security configuration for responding against and mitigating the effects of the attacks. The remediation module brings automation in security orchestration solutions and delivers significant ROI and drive downtime to remediation [90]. The OADS system proposed in [49] has a central controller to implement the established sequence of actions with a process, including exceptions and conditions. Enterprise level security orchestrator [16] has a remediation module that has two main elements: response storage and remediation engine. The remediation engine detects threats patterns. It has a learning logic module that uses machine learning algorithms. Threat data are stored in the response storage once received from the remediation engine. Koyama et al. [51] have discussed the technology to rapidly recover from the effect of cyber-attacks. The proposed remediation module immediately isolates the affected area after detecting attacks based on the information from a detection module and provides recommendations about detected attacks for further analysis and evaluation. These actions of the proposed system are expected to reduce a security operator's burden. The SoSDSec system proposed in [31] incorporates a model layer to manage security policies and security models of organization's assets. The security orchestrator reads and updates policies to achieve automation. SDSec orchestrator is a key element to achieve security orchestration and automation. It works with the communication and coordination subsystem. The communication with security controls is also performed through the orchestrator. The SDSec orchestrator communicates with virtualized functions to coordinate security tasks and thus minimizes management dependencies on security appliances.

*4.3.2.Action Performer.* A controller or action performer controls communication and actions [84]. Security staff can directly control an orchestration platform's various components through a controller [32, 65]. The action performer performs many actions such as send an e-mail to the relevant persons, block an IP address, isolate a virtual machine, trigger a process to initiate a scan and run a script to perform auto-configuration [28]. Poornachandran et al. [32] have referred the data management processing system as security consoles and administrator console that works as a tracking station. The tools enable staff to tackle diverse and ongoing issues [94]. The communication module can be considered as a subcomponent of the controller. The job of the communication module is to bridge between several modules of a security orchestration platform. In addition to maintaining a secure exchange of threat data and policy information, an orchestration platform requires a secure broker or DXL [17, 32, 47]. Elshoush et al. [44] have considered the communication module as a bridge between security solutions and decision-making module. The DXL fabric of [17] provides command and control functions across the entire network. Published, subscribe notification, query response, and push notification are different types of message of DXL layer. Demisto has provided DBot and ChatOps to perform intelligence automation, and collaboration among operations [81].

## 5. Motivation behind Security Orchestration

This section reports the result of RQ2: *"What challenges security orchestration is intended to solve?"* We have identified and analyzed the challenges that promote the practice of security orchestration. Our analysis of the extracted data enables us to identify several challenges as shown in Fig. 8. We have classified the challenges under technical and socio-technical aspects of security orchestration.

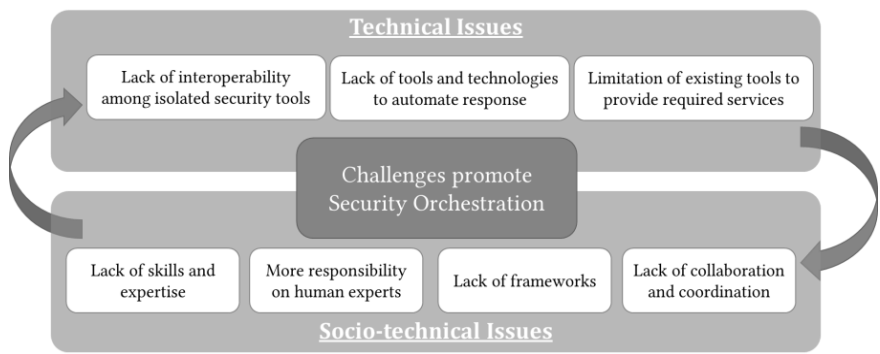

**Fig. 8 Challenges that promote Security Orchestration**

### 5.1. Technical Challenges

Technical challenges are related to technical issues that lead to security problems such as limitations of the IDS to accurately detect intrusion and interact with other security tools, conflict among several security tools in simultaneous run, and dynamic change of tools behavior. Following sub-sections describe the technical challenges that security orchestration intends to solve.

*5.1.1.Lack of Interoperability Among Isolated Security Tools.* Our analysis of the reviewed papers reveals that most medium to large organizations uses several security tools (e.g., IDS, Firewall, and SIEM) to secure their critical data and infrastructure [12, 56, 102, 103, 118]. The main reason ends up installing several types of products is that different vendors provide distinct dimensions of security services and solutions [11, 16, 89, 112, 115]. Moreover, organizations lack a single security tool that can encompass the whole of the security operations. Isolated security tools are considered as a lousy communicator and cannot always assume the presence of another tool [12, 30, 31, 49, 84, 101]. Several security tools fail to guarantee the protection of an organization's infrastructure as they work in an isolated way and focus on solving the specific problem [62, 65, 86]. Several of the reviewed papers have mentioned that it is extremely difficult for a single security solution to detect the distributed and complex behavior of cybersecurity attacks. Moreover, security operators are usually unable to understand their organizational security state through individual security tools working separately [42]. For taking incident response decisions, it is necessary to integrate and analyze the activities of different tools, which are usually designed to work independently and limited by their own services [30, 62, 65]. These tools have their own data representations and interpretation mechanisms. The disparate tools have inconsistent workflow [108], disconnected and non-integrated architecture [12] and lack of standardization for data exchange between different tools [53]. These are some of the reasons that network administrators and security experts find it difficult to appropriately configure and integrate the activities of multivendor security tools that means there is a need of continuous involvement of humans in the entire process of security incident response. The lack of interoperability among security tools results in more responsibility on human experts (briefly explained in section 5.2) and leads to redundant, complex, inefficient incident response process. Existing security management and risk assessment solutions are not designed to collaborate [77]. These solutions do not consider several aspects that affect the evaluation criteria of the threats and vulnerabilities thus make the security procedures incomplete. As a result, with generic security policies, the security management becomes inefficient [77].

*5.1.2. Lack of Tools to Automate Proactive Response.* Our review has revealed that there is a lack of tools to automate the key security activities such as threat intelligence collection and update, alert validation, task investigation, response, and resolution [28, 84, 102, 106, 112, 114]. Organizations need learning tools to automate the manual repetitive tasks. FireEye mentioned security experts spend 95% of their time on the manual execution of repeatable tasks [108]. *AT&T's cyber securities insight* report reveals 90% of their reported cyber-attacks were from known vulnerabilities [112]. Whilst security defender needs to update new threat intelligence quickly, they usually fail to instantly update the threat intelligence [17, 47], promptly update software patches to remove vulnerabilities [66], keep every security tools up to date [102], and enforce policies as soon as they are agreed upon [33]. For a large network, it is time-consuming to update hosts from different vendors that leave the system open for intruder [66, 84, 86]. Ntouskas et al. [77] proposed lack of automated collaborative tools to embed security standards, methodologies, tools, and guidelines to train a security management team as one of the key reasons' organizations lag behind in fulfilling their security needs. Fujitsu emphasizes, an efficient SOC requires automation of the process of threat defense life-cycle [75] to help free up security analysts' time and keep the system up to date.

*5.1.3. Limitation of Existing Tools to Provide Required Services.* Several of the reviewed studies have mentioned that the existing security tools are unable to give full protection to organizations infrastructure [28, 97, 102, 113, 115]. According to *Verizon's 2017 Data Breach Investigation Report* [3], 43% of data breaches utilizes phishing, and it becomes clear that trying to prevent every attack is like playing whack-a-mole. The single standalone detection engine also fails to provide complete visibility of network infrastructure to security staff. In most cases, the detection system generates a large number of false alerts that require extensive analysis [12, 13, 18, 84, 108]. Security experts are overwhelmed with alerts and spend more time investigating and validating false and repetitive alerts. In 2015, Hewlett-Packard reported 48% of their recorded cyber-attacks were from known vulnerabilities that are five or four years old [112]. Organizations need tools that can learn from experts' behavior. There is a lack of platform where security staff can easily integrate security tools, network infrastructure and gather complete visibility of their security systems [28, 89, 93]. Weilinger et al. [68] have reported that IT security tools used by security practitioners fail to address the complexity of their interactions. According to Demisto [114], the inappropriate interface between technology and personnel is the reason security personnel being ineffective and inefficient. The IT security tools provide insufficient support for collaboration, coordination, and cooperation among security practitioners and stakeholders.

## 5.2. Socio-Technical Challenges

Socio-technical challenges are related to the organizational process, policies and rules with respect to cybersecurity. Socio-technical aspects of security in an organization include matters involving business process, skill, resource management, policies, law enforcement and interaction of people with the technical system. Many of the challenges faced by the security community are socio-technical rather than technical.

Socio-technical challenges are difficult to project as it involves interactions between individuals, groups and technical systems. Our analysis of the extracted data indicates some of the key socio-technical challenges that organizations face while handling the security incident. These challenges work as the primary drivers of security orchestration.

*5.2.1. More Responsibility and Workload on Human Experts.* Cybersecurity staff are entrusted with several types of responsibilities that include analysing and dealing with sophisticated attacks [51], manual consultation and writing custom codes to validate alerts through threat intelligence [13, 75, 107], manual extraction of key attributes from threat intelligence data and linking it with relevant data [64], evaluating alerts, correlating data and coordinating the appropriate responses [11] and investigating results [66, 115]. Several papers [12, 13, 30, 31, 51] have reported that the response toward a security incident highly depends on the manual activities performed by security experts. A security expert needs to combine several security tools [68], update threat intelligence, involve multiple administrative systems, include multiple control tools [30], analyze data from new tools [101] and deal with the interaction of inter-component of the modern complex system [21, 56] to perform their tasks. The manual steps are usually the main reason for longer incident respond time [11, 89, 97, 101]. A delay in the security incident analysis happens as a security expert needs to continuously shift between multiple disparate tools to manage different pieces of information from multiple tools [23, 102, 108]. Fujitsu's SOC has considered the manual consultation as one of the most time-consuming steps in incident response process [75].

Hexadite reported [13] that it takes around 45 days for an organization to resolve cyber-attack due to manual response to incidents. According to a set of studies [47, 64, 103, 114], security staff finds difficulties to manually extract features from huge volume of threat intelligence data. Manual configuration, integration of several security tools, implementation and updates are associated with misconfiguration, erroneous response, and policy enforcement [53, 56, 97, 103]. Moreover, security staff face difficulty in manually dealing with the interaction of inter-component of the modern complex system [56]. Several papers [42, 43, 107] have indicated that manually dealing with thousands of alerts to choose the right course of actions result in missing critical attack information.

*5.2.2. Lack of Skills and Expertise.* Security practitioners have reported [11-13, 23, 101, 108, 114] lack of skilled security staff as one of the major reasons for organizational failure to deal with security breaches. Large enterprises are spending billions of dollars on buying and deploying several types of security tools [102, 111] that need up-to-date knowledge and expertise in different aspects of security attacks and countermeasures. Organizations face difficulties to find and retain security staff with the required expertise [23]. Security staffs require a decade to acquire the expertise to fight against sophisticated cybersecurity attacks [97]. It has been highlighted that a global shortage of 1.5 to 2 million cyber security professionals may occur by 2019 [96, 122] and 3.5 million by 2021 [28]. According to CyberSeek, cybersecurity data tools [123], 40,000 jobs for information security analysts remain empty each year in the USA where organization struggle to fill 200,000 other security-related jobs. Organizations have few security experts to deal with thousands of incidents that they are receiving each day [12, 23]. Security staffs need to have an overall knowledge about an organization security policy, network infrastructure, and security tools. One paper reported that organizations are continuously shifting toward modern technology paradigms (e.g., cloud computing, mobile computing, and IoT) that lead to an expanded security attack surface and needs security knowledge, incident response skills and resources for each technology initiatives [106].

*5.2.3. Lack of Regulation and Policy Framework.* One of the major challenges in an organizational security is lack of a fully developed framework for conducting IRP [45], performing coordination and collaboration among incident response [61], seamless implementation and deployment of policies [17]. Some of the challenges mentioned in the reviewed papers include failure to provide clear definition of unwanted traffic and network behaviour [49], significant difficulties in providing clear guidelines to deal with new security mechanisms [31, 68, 73], failure in providing appropriate training for security management [77], lack of guidelines for conducting incident response planning [45] and severe challenges in enforcing and managing security policies. All these types of challenges result in security staff failing in taking a proactive decision against cybersecurity attacks.

*5.2.4. Lack of Coordination and Collaboration Among Stakeholders and Security Teams.* Coordination and collaboration among security staff are important to analyze complex threat behavior. Most cybersecurity communities lack collaborative processes for information sharing. Several papers [20, 45, 46, 48, 55, 68, 78] have highlighted the requirements for having a combined knowledge and experience from several domain experts due to the complexity of network flow and log data analysis. Most of the incident response teams follow no collaborative process while planning how to respond to a particular incident which results in poor strategies plan [45]. Several papers [30, 59, 78] reveal that stakeholders from different organizations are unwilling to share threat intelligence with each other. Jeong et al. [78] have reported organizations' fear of losing reputation is one of the reasons for their unwillingness to share their security circumstances with other organizations. Zhao et al. [59], has discussed that many state and federal governments have developed threat information sharing services which are limited to sharing threat intelligence with the central government. The external organization could not get benefit from this kind of threat information sharing. Still, there are no collaborations among different organizations working in the same domain [30].

Table 5 presents the mapping of the benefits that organizations get through the functionality discussed in section 3.2 that aims to solve the challenges discussed above in this section.

**Table 5** Mapping summary of key activities performed by security orchestration with benefits organization gets

| Benefit of Organizations | Articles | Total |
|---|---|---|
| • Efficient and effective incident handling process<br>• Free up expert's time<br>• Minimize the overall complexity of incident response process<br>• Enhance organization protection and defence system | [11-13, 16, 17, 19, 23, 30, 31, 43, 53, 62, 89, 93, 95, 101, 108, 109, 111, 112, 115, 121] | 22 |
| • Keep analyst focus on threats that demand their ability<br>• Reduce human error<br>• Faster decision<br>• Reduce burden on security operator | [11, 13, 17, 19, 30, 51, 83, 93, 102, 110, 111] | 11 |
| • Experts get the insight of several security controls activities<br>• Organization share contextual device data with third-party system<br>• Reduce and mitigate the risk exposure<br>• Faster decision | [12, 17, 30, 42, 45, 46, 48, 54, 55, 63, 81, 101, 108, 110-112] | 16 |
| • Reduce human error<br>• Improve staff capability to incident response.<br>• Provide standardized process<br>• Reduce reliance on human expertise<br>• Simplification of security management task. | [11, 13, 29, 31, 33, 42, 101, 102, 104-106, 108, 109, 115, 121] | 15 |
| • Accelerate Response<br>• Mitigate conflict installation<br>• Reduce conflict configuration<br>• Minimal the effect of attacks on services. | [29, 33, 46-48, 51, 73, 80, 113] | 09 |
| • Maintain process consistency across security program.<br>• Reduces the manual investigation error<br>• Maintain effective communication and strong collaboration among cyber security teams<br>• Simplifies and accelerates alerts investigation<br>• Accelerates Return of Investment (ROI) | [11-13, 16, 17, 29-31, 42, 51, 73, 78, 80, 104, 107, 108, 111, 115] | 18 |
| • Optimize security staff's capability<br>• Reduce cost<br>• Minimize mistake-prone manual process<br>• Accelerate Response | [11, 29, 45, 47, 63, 108, 111, 113, 114, 121] | 10 |
| • Minimizing mistake prone configuration.<br>• Reduce conflict configuration. | [17, 29, 31-33, 52, 59, 60, 62, 65, 106, 111] | 12 |

| Key Functions | Activity Performs |
|---|---|
| Unify Security tools | - Unifies disparate security tools and process<br>- Integrates enterprise security architecture<br>- Connect detection, networks and endpoint security systems<br>- Perform coordination among security tools activities<br>- Unifying intelligence according to vulnerability<br>- Removes the operational silos. |
| Determine Endpoint for human investigation | - Work as a helping hand for security experts<br>- Informs and educates security analyst about threat behavior<br>- Decide when human insight is needed<br>- Define source of information to help expert solve problems |
| Share Contextual Insight (via platform) | - Gathers threat intelligence from various external sources<br>- Extract key features from threat intelligence data<br>- Provide the contextual insight related to alerts or attack to the security analyst<br>- Provide real-time visibility<br>- Context-aware framework<br>- Gather an overview what is happening in various subnetworks within the environment |
| Translate Complex process into streamline workflow | - Allow experts to simplify high-quality workflow integration<br>- Coordinate the flow of data & task by integrating tools & processes into automated workflow<br>- Enable powerful machine to machine automation<br>- Offers data aggregation to provide in-depth awareness about the environment |
| Provide deployment model | - Resolve incident in minutes to determine appropriate and effective course of actions<br>- Determine the proactive response to threats<br>- Initiate additional investigation based on the level of the attack's complexity |
| Determine appropriate course of actions | - Simplify threat response through integration and automation<br>- Decide additional investigation<br>- Limit execution access and privileges only to workflows.<br>- Dynamically insert security functions into the workgroup based on the policies<br>- Engages the security tools to perform complete monitoring of endpoint and correlate their activities<br>- Performs coordination & collaboration among different anomaly detection |
| Automate Repetitive and manual task | - Automate repeatable task<br>- Automate deployment of security functions over the network infrastructure<br>- Repeatable, executable planning process<br>- Eliminates the need for continuous vendor assessment |
| Automate policy enforcement | - Automate policy enforcement and configuration at runtime<br>- Real-time policy enforcement |

## 6. Taxonomy of Security Orchestration

In this section, we have summarized the results that answer to RQ3: *"What types of solutions have been proposed to adopt security orchestration?"*. We have highlighted the key techniques, tools, and strategies used by practitioners and researchers in the realization of security orchestration. Most of the reviewed studies have proposed platform-based architecture as a strategy for incorporating security solutions to support their integration, orchestration, and automation [12, 13, 80]. McAfee has focused on four engineering approaches to automate the entire threat defense life-cycle - partnership centric, platform-based approaches, reinvented experiences and cloud-centric [12]. All these four approaches are integrated into a single platform to take the benefits of each. We consider the platform-based approach is the core engineering strategy. The orchestration platform is designed to automate various activities of threat defense life cycle. This review has enabled us to propose a taxonomy of security orchestration to support a systematic comparison and analysis of the existing security orchestration solutions as depicted in Fig. 9. The proposed taxonomy consists of several dimensions and sub-dimension for classifying security orchestration techniques.

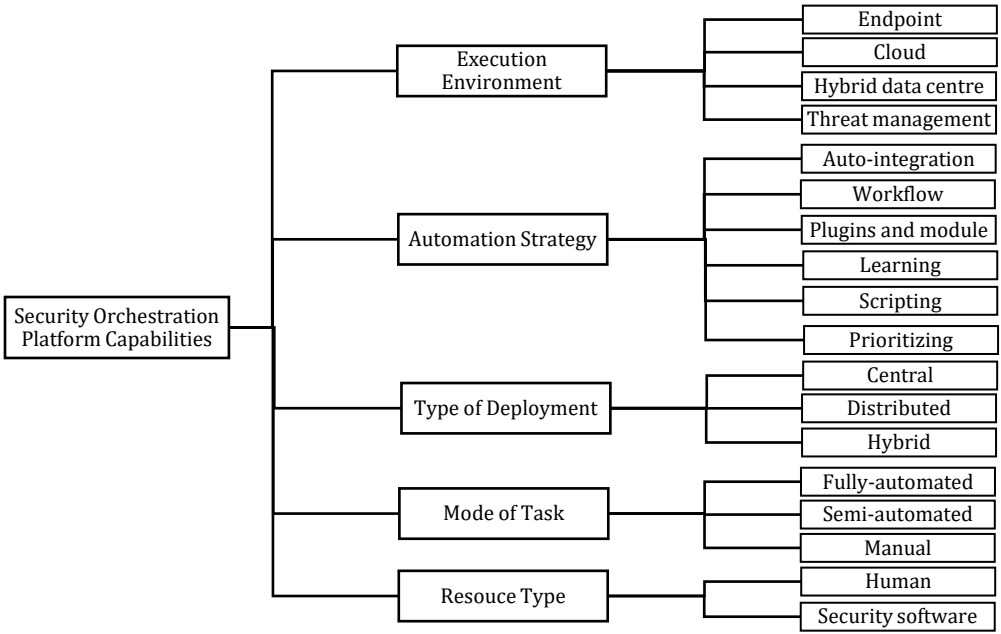

**Fig. 9** A Taxonomy of Orchestration Platform

## 6.1. Execution Environment

To help speedy organizational responses to security incidents with fewer resources, a security orchestration platform's execution environment can be supported by four types of technological solutions that are expected to work together to solve the security issues and challenges; for example, a combination of cloud-delivered data security system and endpoint security for Infrastructure as a service multiple vendor with multi-tenancy features, these solutions need to be considered right in a virtual context. For example, ETSI has proposed a security orchestrator for a hybrid network consisting of a physical network and virtualized network [52]. In the following subsection, we discuss the four execution environments for security orchestration solutions.

*6.1.1. Endpoint.* Most of the organizations have several siloed security solutions on their endpoints. Installing several security solutions in each endpoint and managing endpoint in a large IT infrastructure are becoming challenging and inefficient [92]. An organization needs to monitor, assess, and control all the endpoint devices connected with organizations' network to provide end to end threat protection [89, 92, 95]. McAfee has considered endpoint security architecture from which an organization can expect agent consolidation. In this review, we consider any autonomous entity or software program that can perform actions as an agent. A security orchestration platform can deliver consolidation at the endpoint which can even be the entire portfolio depedning on time [12]. Security orchestration agent can reside in various endpoints storage (RAM, HDD, or SSD) [32].

HEXADITE has proposed security orchestration and automation solutions AIRS that helps an enterprise to connect detection, networks and endpoint security systems. Though the proposed platform seems to be agentless, it uses a non-persistent agent that inject dissolvable probe on endpoints during investigation [13]. Similar to Hexadite, Demisto [115] proposed an architecture that consists of dissolve agent for data collection from endpoint. The workflow tool CounterACT proposed by ForeScout uses multiple agentless discovery methods and integration techniques. CounterACT employs a combination of active and passive discovery methods to classify organizational devices based on the network [111]. Without installing any software agent or enrolling any management unit to a device, it first connects the device to the network. This reduces the overhead of an administrator to check each device and manually assign policies to each endpoint. The resilience engine proposed by NTT controls multiple devices at appropriate points according to the type of attacks to isolate the affected regions [51]. The security orchestration framework also supports distributed endpoint with DXL over an enterprise network [17]. DXL is built on top of Enterprise

Service Bus (ESB) technology and provide an abstraction layer between different types of connected endpoint devices. Through a security orchestration platform organization can provide constant protection irrespective of where an endpoint device is located.

*6.1.2. Private and Public Cloud.* Cloud computing and its related technologies have created the need for new generation security technologies. An orchestration platform can be built as a single integrated solution to provide cloud-delivered data security [12, 92]. The motive of cloud platform is to build software as a service with the required levels of performance and availability. An organization can easily integrate their security module into a cloud [92]. Example of such services includes web protection, sandboxing, security broker, data loss prevention and encryption. A cloud security platform continues to support next-generation platform that is built beyond VMware and Amazon web services to add Azure, OpenStack, Docker containers and emerging services. McAfee has proposed McAfee cloud ePO software to support consolidated management across their cloud management technology [12]. The work proposed in [31] is designed to deal with heterogeneous cloud environment and automated security operation in Software Defined Infrastructure environment. The proposed solutions handle VM movement over dynamic infrastructure and provide transparent security management facilities. The resilience security technology for rapid recovery from cyber-attacks also works for network services in cloud environments [51]. The enterprise-level security orchestrator installs a mirror of the security orchestration in a cloud [16]. They have proposed security orchestration engine for server and client and the orchestration solutions can be used in public, private or even external cloud [16]. Using cloud as an execution environment helps an organization to have a scalable, flexible and adaptable infrastructure.

*6.1.3. Hybrid Data Center.* The evolvement of a data center to SDDC has created new security-related challenges for organizations. Additionally, the increasing trend of distributing more workload on data center and public cloud has also increased security challenges. The security orchestration system and resilience engine proposed by NTT are mainly designed for a data center [30, 51]. McAfee aims to build an integrated security system to deliver visibility and security to a cloud-enabled data center [12]. McAfee's security orchestration platform includes global load balancing infrastructure with Content Delivery Network (CDN)/ peering data center [12]. Intel has introduced the open security controller, a security orchestration platform to orchestrate virtual security in SDDC [29]. The purpose of this platform is to make security management visible, effective, agile and scalable by providing automated, dynamic and synchronized security services for software-defined infrastructure. It provides seamless brokering services between SDN and VNF. It is optimized for OpenStack and VMware cloud environment. A security orchestration platform gives visibility across network and server tiers and public/private cloud data centers [92]. Dynamic micro-segmentation is performed for private clouds and workload auto-discovery is done for public clouds [12, 29]. The concept of micro-segmentation restricts the access and tailor security configurations. This gives better threat protection and faster remediation than siloed approaches. Security orchestration helps security administrator to span their security model from organizational data center [92].

*6.1.4. Threat Management.* A SOC suffers from a large volume of data, events and Indicator of IOC to prioritize true attack in process or in the golden hour of post-breach [44, 46, 47]. A security orchestration platform helps in analyzing the security threat data by providing security analysis, threat and vulnerability management, attack detection, attack investigation and streamlines incident response [12, 30, 51, 59, 92, 95]. Threat management includes prioritize threats in progress and also in the golden hour of post breaches [12]. It helps security analytics to continue advanced data management, risk assessment, correlation and deal with both volume of security data and increasing sophistication of analysis [49, 53, 54, 59]. It automatically investigates attacks during and after a breach. It also provides both on-premise and cloud-based analysis. Threat Connect has proposed an intelligence-driven platform to manage both internal and external threat data and turn them into actinal threat intelligence [86]. Security orchestration provides a central place for data aggregation, analysis and enrichment of security threat data. Security orchestration allows providing technologies like attack reconstruction that helps an organization to identify and response at a full attack level not just at an event or malware level. An organization can use a security orchestration platform to centrally manage the threat and automate entire life cycle of threat defense.

## 6.2. Security Automation Strategy

Our review has revealed that a security orchestration platform uses a combination of several types of automation strategies. Orchestration of different automated steps is needed for effective incident responses that suit all types of organizational activities that are integration, aggregation of data, auto investigation or analysis, finding proper course of actions and deciding remediation process. The security automation realization approaches concern specific methods/tools. HEXADITE has highlighted five distinct approaches to security automation adopted by current vendors – workflow tools, orchestration tools, scripting tools, prioritization tools and intelligence security automation [104]. From the analysis of the reviewed literature, we consider the intelligence security automation as security orchestration platform that includes some of the available automation tools to orchestrate and automate the incident response process. Demisto has mentioned automation and human tasks need to be interweaved and worked together in a seamless fashion to achieve the desirable goal [115]. We have outlined the automation strategies that are used by practitioners in various organizations.

*6.2.1. Auto Integration.* We have placed the connecting or integration tools that are used to automatically connect existing security tools through APIs to streamline an incident response process Under this category. Whilst some practitioners have mentioned the connecting tools as orchestration tools [104], this is not the only purpose of using a security orchestration platform. A security orchestration platform has tools to automatically connect and integrate a full stack of security systems [11, 80, 104]. Several reviewed papers have also proposed to connect organizational hardware, software, and control unit into a security orchestration platform [113, 121]. This work as a layer of connective fabric that makes the tools work together. The integration tools help isolated security tools to interoperate with each other. Organizations can easily buy a new point of product as the integration tools automatically connect and integrate new systems into existing one and make the necessary changes in a system. An organization can dynamically insert security functions into any workgroup based on their policies [29]. The ControlFabric interface by ForeScout uses an open standard based API to perform bi-direction integration [101]. Building a fully autonomous integrated set of tools is a very difficult due to the heterogeneous nature of multivendor security solutions. The work reported in [17] allows the integration of a third party software. Security management software can connect to the orchestration framework by connecting to DXL. It provides all the command control functions across the entire network. The DXL also provides API embedded with McAfee agent. However, if a security orchestration process is not well-defined [104], there are not many benefits in just connecting the existing security tools. An orchestration platform must have some well-designed framework and workflow to perform the required actions.

*6.2.2. Workflow.* Organizations usually use workflow tools to streamline an incident response flow and communication. Workflow tools are depicted as a solution to gather and enhance alerts that automatically send instructions to analyst, auditor, and other security systems [11, 29, 90, 101, 104, 115]. Some workflow tools provide a standard framework specifying user roles and types of actions need to take for certain types of alerts [28, 84, 86, 90, 95, 104]. FireEye has built a security orchestrator to design a workflow [80]. A security orchestrator can help organizations to better organize incident response flows with a built-in ticketing system. According to HEXADITE and KOMAND, the workflow tools automate data gathering and communication processes, leaving the investigation and remediation actions for security staff [11, 104]. Security team creates a sequence of automated tasks to perform the tasks in a logical sequence with a chained data flow [19, 21, 84, 86, 115]. Demisto has mentioned designing the workflow for automation of playbook to weave human analyst into the middle of these workflows and playbooks. Though some practitioners have designed the workflow in such a way that it will also trigger investigation and remediation actions [29, 84]. Some workflow tools like the one proposed by McAfee, drive cross endpoint workflow and built natively into an endpoint security architecture [12]. ForeScout has built *CounterACT* that uses a rule engine and a workflow engine to automate the workflow for instant decision making to deal with security incidents [101]. This helps an organization to automate the security process across mobility management and endpoint platform. Invotas Inc. has designed a multistage workflow which includes the workflow of automatically connecting different security tools [121]. Majority of the security orchestration practitioners have built the workflow based on use case scenarios [11, 13, 80, 90, 101, 114]. An organization can define its strategies about how to respond to certain security events. The cybersecurity playbooks keep a record of this in the form of a workflow rule that an orchestrator uses to autonomously control attacks [23, 31, 84, 90]. Workflow tools do not enable the integration of heterogeneous inter-organizational information and security systems.

*6.2.3. Scripting.* Scripting tools perform actions based on custom code written by security staff, who use the scripting tools to configure existing playbook, security tools and policies. An organization requires skilled developers to consistently write and maintain code by performing in-depth investigation [23, 90, 104]. Scripting tools can be considered as an execution engine that executes the script or configurable code. The security orchestration system proposed in [30, 51] uses scenario-based autonomous control of multiple virtual appliances to implement security measures. A security operator can implement the measures regardless of their skill level. Scripting capabilities also include writing custom workflow and integration [90]. An organization with resources to investigate and remediate threats can use scripting tools to perform automation. An organization needs to have both budgets and resources for using scripting tools. Defining new policies and designing scripts according to organizations' budgets and resources can be considered under this category. The policy maker explicitly writes code to reconfigure some parts of a network. The enterprise-level security orchestration has an orchestrator routine to make call to safeguard software packages via automated interfaces provided by safeguard interface modules [16].

*6.2.4. Prioritization.* Prioritization tools help security team to decide critical security alerts. These types of tools normally assign a score to alerts to reflect more critical and urgent alerts to prioritize security events [21, 43, 66, 75, 86, 104]. Most organizations have some sort of prioritization tools with their detection system that automatically investigates and correlate alerts to reduce false assumptions and give experts a list of critical alerts, which are produced by organizational detection system. Major data breaches show that in most cases organizational security teams missed critical alerts. SIEM [43, 84, 114] is a popular prioritization tool that collects and aggregates alerts from different security tools and prioritizes true alerts and discards false alerts. Tayeb et al. [43] have proposed an ontological reasoning approaches to reduce the false alerts by correlating alerts. Some of the orchestration platforms have used the existing SIEM technology to prioritize the alerts.

*6.2.5. Learning.* A set of studies have used Artificial Intelligence (AI) technique and game theory model to make security system intelligent [21, 44, 48, 66, 73, 77]. McAfee has proposed to expand the security orchestration platform capabilities by including behavioral security; for example, pre-execution, post-execution, machine learning and more [12]. Several of the reviewed studies [12, 21, 79] use machine learning based solutions to analyze security behavior. AIRS platform, an orchestration platform, proposed by HEXADITE uses AI to automate the activities of several security tools [107]. HEXADITE has proposed the used of AI as a critical capability for automated security technology [107]. Without automatically learning, it is not possible for any security orchestration platform to predict uncertainties. Demisto introduces ChatBot – a learning tool to combine intelligent automation with collaborative, human social learning and experience [114]. For certain threat behavior, defining rules and designing workflow work fine. With a world full of uncertainty, an orchestration platform must be able to learn from a security expert's behavior and threat data. A combination of AI techniques (such as Machine Learning and Genetic Algorithm) has been used in security orchestration platforms' automated learning modules. Seclius [42] uses a set of instruments to learn from the dependency of systems' assets and captures information flow between file and process. The authors have also developed an algorithm to use with the set of instruments [42]. As a result, an administrator does not need to define the low-level input.

*6.2.6. Plugin and Module.* For this category, we have classified small programs or software that organizations can independently select and install based on the required configuration. A security orchestration platform can integrate plug-ins to automate various activities and create workflow [65, 90]. Siemplify has introduced a plugin framework for security orchestration that makes security tools accessible and easy to integrate into incident response workflow and automation [90]. Komand has introduced several plugins to include in an organization's environment [11]. Each plugin has a set of tasks for a specific set of activities. FireEye orchestrator also uses predefined plugins to perform workflow integration [80]. This makes a security team more agile. Module based automation strategy helps an organization to choose an integration module based on organizational infrastructure, policies, and configuration. ForeScout Module supports more than 70 third party solutions to automate various activities of security tools [101]. ForeScout open integration module allows customers, system integrators and third party product vendors to integrate their products with ForeScout's CounterACT and communicate with each other. Both module and plug-in add specific features to an orchestration platform that is why we have placed them under a single category.

## 6.3. Deployment Model

A security orchestration platform includes several types of components. The platform might have several structures to manage its components and activities. Some of them form a distributed structure, while others become part of a central management site for a large-scale deployment [65]. ForeScout has mentioned three deployment models for their proposed orchestration and automation platform: centralized, distributed, and hybrid deployment architecture [111].

*6.3.1. Centralized Deployment.* In a centralized deployment architecture, an organization has a centralize orchestration manager to communicate, manage and deploy policies to multiple orchestration appliances in a data center or major sites. Several papers [19, 65, 86, 89, 90, 93, 118] have indicated that a centralized security orchestration solution is needed to provide security operation central team better understanding of the state of security throughout an organization for faster and efficient incident response actions. For example, NetSec has proposed a central management for a large-scale deployment of orchestration platform [65]. Whereas, ThreatConnect [86] has proposed a central intelligence driven platform to manage threat data in a single place. Centralized management configuration is necessary for optimal security enforcement. In this type of deployment, the appliances need IP connectivity to remote sites in order to manage devices and other endpoints located there. Traffic from the remote location is sent to a centralized orchestration platform via a predefined interface for monitoring and assessment. A security orchestration appliance can monitor the activities of the user directory, DNS and DHCP to detect threats or potential rogue activity and initial remediation [111]. An organization manager contains the database of endpoints (active or passive) from the appliances it manages [64].

*6.3.2. Distributed Deployment.* An organization can use a decentralized deployment model for utilizing a mixture of security orchestration component located in both central facility and various remote sites [59]. A security orchestration controller manages and controls the activities, provides policies to orchestration service consumers, physical or virtual and maintains a database of active and inactive endpoints. A distributed deployment enables the use of virtual firewalls, virtual security services, browser redirection, and endpoint authentication to a server when a local orchestration platform is at that site. A distributed organization, large data center, cloud platform and large IT infrastructure require distributed deployment of security policies and protocol that can be achieved by distributed security orchestration appliances over multiple endpoints. Incorporating distributed security analysis and monitoring allow an organization to deliver tighter security policies and better protection against emerging cyber threats [92]. ForeScout [111] has introduced orchestration organization controller functions that are the central notification points, where the communication occurs via email or syslog and bi-directional SIEM services via CEF or LEEF messaging to perform endpoints actions and to notify systems to endpoint status. Radwane et al. [46] propose a distributed collaborative architecture to perform cooperation and placement of defense entities on organizational systems to defend against DDoS attacks. They have utilized the concept of distributed hash table and overlay network to perform the distribution and placement of security solutions. Fung et al. [48] have used Chord overlay network to implement the protocol of their distributed system.

*6.3.3. Hybrid Deployment.* The hybrid deployment model uses a mixture of security orchestration platform appliances in a central location and at remote sites. The orchestration controller maintains a database of the infrastructure and issues policies for the appliances. Chen et al. have proposed a centralized controller for managing the distribution of appliances [64]. A hybrid deployment implementation supports virtual firewalls, browser redirection, and authentication verification of an endpoint to a server when a local security orchestration appliance is deployed at that site [111]. Elshoush et al. [44] have highlighted several hybrid deployment models for a collaborative intrusion detection system. Their proposed architecture can also be considered a hybrid architecture. The security orchestration framework reported in [17] performs distributed sensing over server and client. The proposed system performs a centralized aggregation of data and enforcement policies both on the server and clients. In [17], a centralized server has the visibility to see the entire context and communication layer DXL that is highly scalable based on an elastic architecture that supports multiple deployment options. Multiple controls can be connected and deployed at diverse locations with distinct types of control and visibility where a data exchange layer (DXL) provides a fabric to help them operate as a unified system (super-control).

## 6.4. Mode of Task

A security orchestration platform generates the remediation actions that are both automated and semi-automated [80, 104]. Some actions need human involvement which depends on organizational policies and rules. Security orchestration requires a combination of machine drive and human-led process workflow to optimize security operations [90, 115]. Actions can be triggered by an analyst or triggered when a new artifact is added to an incident [115]. As the security orchestration platform works as an intelligence assistance for the security experts, who should conduct automation selectively based on their resources and needs. The incident response can be fully automated or semi-automated [84, 90] depending on the nature of the tasks to be carried. For example, a task such as notifying stakeholders, assigning incidents and enriching data with context can be automated safely, but the actual containment of data breach, analysis of unknown threats frequently require human in the loop [84]. The online evaluation framework proposed in [42] does not automatically respond to an attack. Instead, it is designed to help security administrators by providing situational awareness capability. Xiaojing et al. [47] have made the feature extraction and analysis of threat intelligence data fully automated. They have proposed a fully automated cyber threat intelligence gathering solutions to lessen the manual job of threat intelligence analysis.

A security orchestration system based on global threat intelligence platform provides both fully automated and semi-automated tasks [51] that help to automatically classify the detected cybersecurity attack, investigate whether or not the available countermeasures are possible. The system also investigates the possibilities of automated response (automatic generation and notification of response recommendations that guide the decisions of a security staff). A platform allows users to choose their level of security and types of responses. A user can take control of the system to combine various security sensors and appliances. Similar to these the system in [31], an administrator specifies the service requirements based on security controls and needs.

### 6.5. Resource Type

The analysis of the reviewed material reveals that the functionality and performance of security orchestration depend on human expertise and security software of an organization. We consider organizations' security software and human resources as two most important resources of a security orchestration platform. Building a security orchestration platform on top of a clumsy list of security software that is supported by unskilled security team will not bring many benefits to organizations.

*6.5.1. Security Software Resource.* In this category, we consider the existing security solutions provided by third party vendors or owned by an organization. Most of the security orchestration platforms assume that organizations already own multivendor security tools. Several papers have mentioned a range of security tools while designing orchestration platform for small to the medium organizations [13, 17, 29, 49, 60, 66, 77, 101, 108]. Some key types of security tools used by organizations are SIEM, forensics tools, signature-based control tools, firewall, IDS/IPS, anti-malware, antivirus, perimeter security tools, ticketing solutions, traffic inspection tools, compliance tools and vulnerability scanners [43, 108, 115]. McAfee has classified the existing security tools into attack detection and attack investigation [12]. Komand has mentioned IDSs, firewalls, ticketing tools, and team communication tools as the minimum numbers of tools an organization must have to build a security orchestration platform [11]. To further enhance the performance, Komand has considered threat intelligence, malware analysis tools, and forensics tools as the next layer of tools [11]. Komand has also considered some additional tools such as applications for vulnerability scanning, phishing investigations, threat hunting, monitoring tools, and malware protection tools [11]. Most of the reviewed literature has considered anomaly detector to analyze traffic for identification of potential attacks and abnormal traffic. Kenaza et al. [43] have performed cooperation among IDS, network scanner and vulnerability scanner to reduce alerts volume. Feitosa et al. [49] have mentioned two types of anomaly detectors: hardware and software based. They have mentioned several hardware tools to capture network traffic [49]. These tools can also inspect network traffic in real time. Several tools, techniques, and systems are used as software-based anomaly detectors such as, IDS (Snort, BrO, & Prelude), Honeypots (Honeyd, & Nepenthes), and open software prototypes [31, 49]. The software security units give alerts to an orchestrator and receive scripts command from an orchestrator.

*6.5.2. Human Resource.* Human is an essential part of a security orchestration platform. Security analysts, security engineers, forensic experts, network administrators, security administrators, director of security operation center including security orchestration designers and security orchestration and automation engineers are considered as human resources for a security orchestration platform [20, 44, 61, 65, 68, 73-75, 77, 98]. Demisto has considered any security staff who perform day to day security operation as a human resource [114]. An organization must have experts to assess organizations' security infrastructure. According to a report by NSSLab, [20], a security architecture can ensure the organization security for an assigned level across the entire threat defence life cycle by assessing organizations existing security infrastructure. Before setting up orchestration operation, the security orchestration designers need to communicate and work closely with security analyst to make sure that the orchestrated process is well-understood [23, 90, 94, 96]. Security experts are the one who perform coordination, timing, moderation, prioritization and enforcement algorithm for policies based on organization requirements [20]. Human resources must be able to fully leverage the power of a security orchestration solution [90].

A human-centric security orchestration model is necessary where the dashboard and planning tools will make automation work. A security orchestration platform is built to work as an intelligence assistance of a security expert. According to Bruce Schneier, automation is only possible in the environment of strong certainty, where everything is related to the planning of certain actions and synchronization of activities [14]. On the other hand, the uncertain world needs direct execution, initiative, and prioritized command. He emphasizes that it is not possible to replace humans, rather, humans are required in security orchestration to make the machine intelligence effective for security response actions. Zonouz et al. [42] have mentioned an online evaluation framework to help administrators. For a coordination model, the work reported in [77] has proposed four groups of users where they have considered security and business continuity team as a group and administrator as another group. The other two groups are a group of local users and group of external or corporate users [77]. Security teams vary by size, vertical and expertise and what an organization needs from threat intelligence [86]. A security orchestration platform should be designed in a way that can work with all sizes, maturity levels, and groups of a security team.

## 7. Discussion

This review has initially introduced and analyzed the relevant aspects that motivate the need of security orchestration. Throughout this review, we have identified and categorized existing security orchestration solutions. There is an increasing realization that security orchestration platforms can enable significant progress towards achieving the goal of security as a service/ utility. Over the years, several technologies such as SIEM and Distributed Intrusion Detection System have been proposed as solutions to the challenge of providing security as a service. However, security orchestration is still in its early stage of development that has a significant potential for research and innovation. One of the areas of research is standardization as most of the security vendors are coming up with their own orchestration platform solutions that have proprietary interfaces or plugins to integrate and access different security tools and services. This heterogeneity works as one of the major barriers to large-scale implementation and realization of security orchestration. Hence, security orchestration solutions require new levels of collaboration and performance. The solutions also need to be adaptive to organizational structures that are quite dynamic in these days. According to a report of Research and Markets [124], by 2021, the security orchestration market price will hit 1.6 billion USD. A security orchestration platform needs to engage security staff fast enough to make a significant difference in the response time. Security orchestration platform needs significant amount of research to create results for immediate incident response application to unforeseen cybersecurity events. This review has enabled us to assert that large scale empirical studies of the security orchestration platforms and practices in the "Wild" will greatly benefit the efforts aimed at addressing the obstacles to security orchestration in different organizational settings.

### 7.1. Open Issues in Security Orchestration

This literature review constitutes a first step towards reaching a common consensus as we examined several state-of-the-art and state-of-the-practice security orchestration platforms and compared them with the existing literature. Our review has revealed that the existing security orchestration solutions suffer from several open issues. We have analyzed the open issues from three key aspects of security orchestration – people, process, and technology as shown in Fig. 10.

- Security orchestration is mainly aimed at increasing automation of the security related tasks that primarily rely on human expertise. The human need to be involved in the loop of orchestration and automation. With automation, security orchestration requires experts who can easily take the benefit of the automated decision and take the control when automation is inappropriate. There needs to be significant collaboration among different level of staffs involved in dealing with the security orchestration processes and technologies as each team may have different responsibilities, priorities, and metrics. Whilst security orchestration automation efforts are based on scenarios known to security practitioners, security vendors and organisations need to develop and deploy more formal workflows and playbooks for security orchestration. Experts involved in the process of orchestration and automation require proper training to gain a common understanding of the workflow, tools and techniques. An organization requires a security architect who can ensure the involvement of risk management and guidance to managed policies. Though one of the motivations behind security orchestration is to handle the collaboration among stakeholders and security experts, the security orchestration platform itself requires strong collaboration among business risk owners, risk assessment team, security operation center and IT infrastructure manager. The analysis of our review has identified that the current security industry lacks training related to secure practices. That means the organizations and security community both need to train the current and future staffs to keep pace with the wide adoption of security orchestration platforms and conceptualize the data needed to acquire the insight into security events.

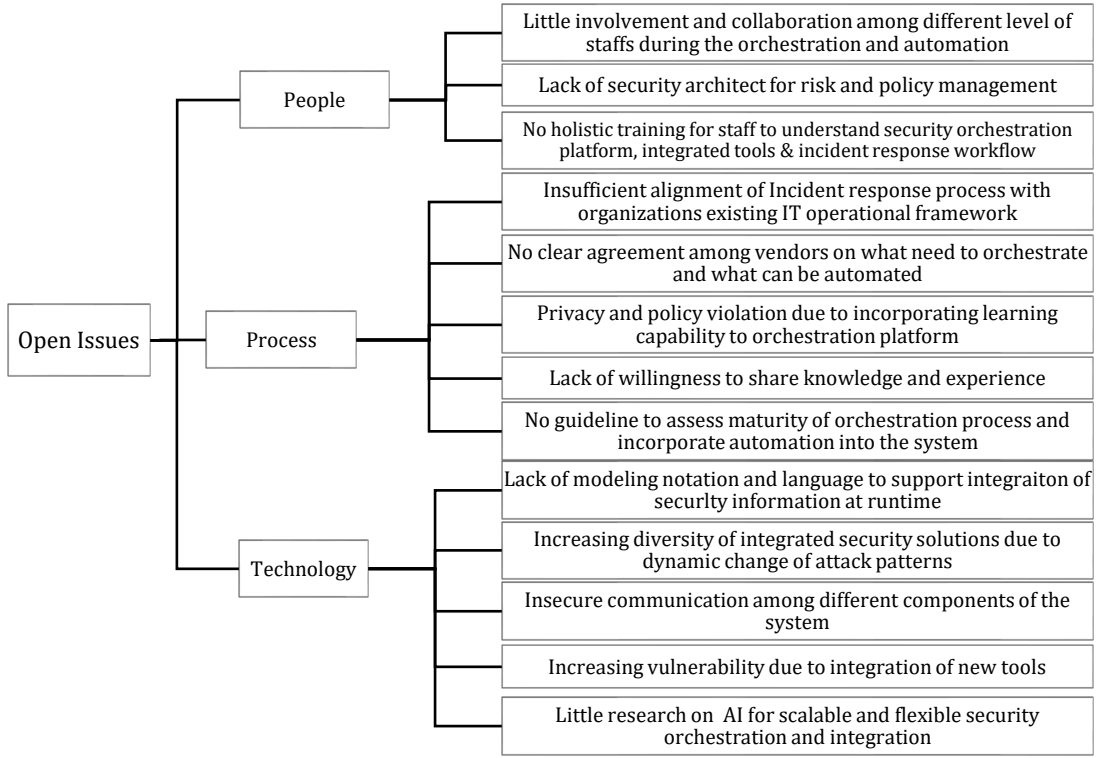

**Fig. 10** Open issues in security orchestration platform

- The incident response process must be aligned with an organization's existing IT operation framework. The organization needs to have a clear idea of what they can automate and what they need to orchestrate. Hence, there is no agreement on what to automate and what to orchestrate. Nevertheless, some research also refers to orchestration but does not specify its meaning, they often focus on building plugins to integrate existing solutions. Security orchestration platform needs to access organization policies and other security solutions data to make relevant decisions. Whilst security staff is empowered to streamline incidents including addressing the issues raised by an orchestration platform, most organizations do not share their threat intelligence with others. This situation can lead to trust issues among organizations.
- Organizations need to assess their respective maturity in the security orchestration and automation [84]. There is still no guidance about how to assess an orchestration process maturity and know when and how to incorporate automation into its systems.
- Whilst there is an increasing recognition of the importance of security orchestration and automation, the practice of security orchestration is unbalanced. Technology should reach a level to further support the development of an agreement on the definition of orchestration and automation in cybersecurity space. With the advancement of technologies, new vulnerabilities are found and exploited every day. The dynamic change of attack patterns causes the increase of diverse security solutions. The orchestration platform should be adaptable with emerging technologies. As stated in the previous sections, orchestration in cybersecurity constitutes an interdisciplinary research area that adopts concepts from research in cybersecurity solutions, SIEM, cooperative IDS, distributed IDS, orchestrated and automated incident response. There is a need of significant research on the modeling notations and languages to support the integration of security-relevant information into streamlining incident response workflow at runtime. There has been no systematic approach to provide a standard API to perform the integration and handle the communication among different components of an orchestration platform. There are a very few orchestration platforms that provide plugins or support for all the existing multi-vendor security solutions. One key challenge is to secure the integration and communication of security systems. Another area of future research and development is the application of AI in security automation technology, which can extend, and/or replace where possible, human cognitive processes for security decisions. The existing orchestration platforms are not scalable and flexible enough to handle the heterogeneity of security team structure, sizes, and expertise levels.

## 7.2. Reference Architecture for Security Orchestration

Security orchestration is an emerging area of research and practice. There is not much accumulated knowledge and experience available to support industrial decisions making for different aspects of security processes and tools for security orchestration. Siemplify [118] suggested the importance to have a delicate balance between human intervention and automation. New security solutions are expected to be adaptable to the existing orchestration platforms. However, a centralized platform for security orchestration and automation usually incurs huge overhead cost and can be a single point of failure. Due to the ubiquitous realization of cloud, edge, fog, and mobile computing, there needs to make suitable changes in the ways of deploying security orchestration platforms, which consider certain properties such as context specific, knowledge sharing, self-reinforcing and dynamicity. The services provided by security orchestration and automation should be fragmented into siloed. It should perform multiple actions in parallel to unleashed the best results out of it and take action against a threat without delay. This requires choosing appropriate architecture, for example, microservice, service-oriented, monolithic, and so forth.

A Reference Architecture (RA) for security orchestration will help define a model to characterize different components of a security orchestration platform and the relationship among different components of the platform. The RA can provide an overview of the flow of communication underpins a workflow. An orchestration platform might have a distribution layer which needs to consider the response time, redundancy and accuracy. An orchestration platform needs a well-designed and rigorously evaluated architecture that can support easy integration and smooth interoperability of components and tools developed for various domains by different vendors. There needs to be architecture level support for visibility and comprehensibility of the functioning and interactions of different components of an orchestration platform that should operate transparently. An orchestration platform's architecture is expected to be dynamically adaptable to the changing security vectors. It is a significant research challenge to design and evaluate a suitable reference architecture for a large-scale realization and materialization of security orchestration platforms.

The MLR has identified the essential components of a security orchestration platform that is expected to have certain quality attributes such as scalability and flexibility. We can conclude that there is an urgent need of conducting research for identifying and leveraging suitable styles/patterns in designing and evaluating architectures for security orchestration solutions.

## 7.3. Limitations of this Research

This study has some potential limitations. Since security orchestration is an emerging paradigm with mixed and inconsistent terms, the search string used to identify the relevant papers may not have included some words that might be used for security orchestration. The inclusion and exclusion criteria used to assess and select the reviewed studies have been defined by the research team. The focus of this review does not include an in-depth discussion of the limitations of the reported solutions. We encourage a reader to take the above-mentioned limitations into consideration while using the findings from this paper. Moreover, some organizations' security orchestration requirements may not fully be met by any of the reported security orchestration technologies.

## 8. Conclusion and Future Work

Security experts usually get overwhelmed by the task of monitoring and handling an increasingly huge pool of security alerts generated by a diverse set of security tools. Hence, they may fail to act in a timely manner to deal with security incidents due to the manual and repetitive job of receiving and combining security alert information from multi-vendor security tools. Security orchestration is aimed at supporting security staff for effectively and efficiently monitoring and dealing with security incidents by enabling coordination and collaboration among the heterogeneous independent security tools. Integrating and orchestrating various activities of security tools in an organization need a comprehensive view of a security orchestration platform. Recently, all sort of organizations have started taking interest in adopting security orchestration. However, academic research is yet to catch up with the increasing trend of technological innovation and practical adoption of security orchestration. Security tool vendors do not share a common/similar understanding while developing and supporting tools, process, and technologies for security orchestration

We have systematically selected and rigorously analyzed the security orchestration solutions provided by various practitioners and researchers to provide a good understanding of this emerging paradigm. We also intended to explore the challenges and the possible future trends of security orchestration research and practice. Our review has tried to address three research questions: 1) *What is Security Orchestration?* 2) *What challenges security orchestration intend to solve?* 3) *What types of solutions have been proposed?* We have identified and analyzed critical aspects of security orchestration solutions founded in 95 papers, which have been selected based on a pre-designed review protocol. To the best of our knowledge, this MLR can be considered as the first attempt toward systematically reviewing and analysing the literature on security orchestration.

The analysis of the extracted data to answer the RQ1(i.e., *What is Security Orchestration?*) enabled us to explore several definitions of security orchestration provided by practitioners and come up with a working definition for the research on the topic of security orchestration. The definition of security orchestration provided in this paper is expected to help practitioners and researchers interested in this topic. Most of the reviewed literature has considered security orchestration as a platform that integrates and unifies various security tools and activities for prompt response to security incidents. The review has identified the key functional and non-functional requirements that a security orchestration platform. Our analysis of the identified functional requirements has revealed three key areas of focus of security orchestration: *1) unification* - which is to unify security tools activities *2) orchestration* - which relates to the process of translating complex process into streamlined workflow and *3) automation* - which is the process to select suitable course of actions to enable automated incident response. Our review has also identified the key components of a security orchestration.

A security orchestration platform is expected to address several technical and socio-technical challenges for which the review has identified the key techniques, tools, and strategies. We have proposed a taxonomy for a security orchestration platform from five key dimensions: 1) execution environment, 2) automation strategies, 3) deployment type, 4) task mode and 5) resource type; which is further split into sub-dimensions. This taxonomy gives the view of the multidisciplinary nature of a security orchestration platform. An organization can compare several security orchestration solutions using the reported taxonomy, which can provide security practitioners with insights into security orchestration platforms' usability in different but interdependent processes.

Despite the widespread adoption of security orchestration technologies and practices in recent years, several management issues (such as legal issues, trust management, adaptability, scalability, and usability) have mostly been neglected and need to be addressed. This review has identified several future research areas as follows:

- There is a dearth of solid studies aimed at evaluating the usefulness (e.g., effectiveness, usability, and accuracy) of security orchestration tools and techniques. There is an important need of research for defining evaluation criteria and metrics to empirically evaluate different aspects of security orchestration solutions including the promised functional and quality requirements.
- There should more research on designing suitable reference architectures for supporting the activities of security orchestration platforms, whose common and variable layers can be known to security tools developers and integrators who are responsible for integrating a diverse set of security tools into a security orchestration platform. A reference architecture will also help to decide where to automate security processes and where a security orchestration engine is required.
- There is a need of research on the expectations and mechanisms of ensuring privacy of data when it travel among different security tools, connected with a security orchestration platform.
- As the next step, we plan to carry out research on some of the identified areas such as providing a security orchestration platform' reference architecture for guiding how to identify the required components of a security orchestration, the interconnection among the components, and data and control flow among these components. Since security orchestration for the organization is at an early stage of development and adoption, we believe that this review paper will serve as an important reference for future research in this domain.

**Acknowledgment**

The work is partially supported by CSIRO/Data61, Australia.

**Appendix**

**Table 6.** Table of Notation

| Acronym | Abbreviation/ Description | Acronym | Abbreviation/ Description |
|---|---|---|---|
| AI | Artificial Intelligence | MLR | Multi-vocal Literature Review |
| AIRS | Automated Incident Response Solution | NFV | Network Function Virtualization |
| API | Application Programming Interface | NTT | Nippon Telegraph and Telephone |
| DBot | Demisto's chatbot | NVD | National Vulnerability Database |
| DDoS | Distributed Denial of Service | OADS | Orchestration-oriented Anomaly Detection System |
| DHCP | Dynamic Host Configuration Protocol | OVAL | Open Vulnerability and Assessment Language |
| DNS | Domain Name System | ROI | Return on Investment |
| DXL | Data Exchange Layer | RQ | Research Question |
| EC | Exclusion Criteria | SDDC | Software Defined Data Centre |
| ETSI | European Telecommunication Standard Institute | SDN | Software Defined Network |
| IC | Inclusion Criteria | SDSec | Software Defined Security |
| IDS | Intrusion Detection System | SE | Software Engineering |
| IDMEF | Intrusion Detection Message Exchange Format | SIEM | Security Information and Event Management |
| IOC | Indicator of Compromise | SLR | Systematic Literature Review |
| IPS | Intrusion Prevention System | SOSDSec | Service-Oriented Software-Defined Security |
| IRP | Incident Response Plan | STIX | Structured Threat Information Expression |
| IT | Information Technology | TAXII | Trusted Automated Exchange of Intelligence Information |
| LEEF | Log Event Extended Format | PSI | Premise-aware Security Instrumentation |
| MD5 | Message Digest 5 algorithm | VNF | Virtual Network Function |